\documentclass[12pt]{IEEEtran}

\IEEEoverridecommandlockouts                              

\usepackage{textcomp}

\usepackage{amsmath,amssymb}

\usepackage{amsthm}
\usepackage{amsfonts,mathrsfs}
\usepackage{color}
\usepackage{graphicx}
\usepackage{epsfig}
\usepackage{subfigure}
\usepackage[utf8]{inputenc}
\usepackage{amsmath}
\let\labelindent\relax
\usepackage{enumitem}

\usepackage{algorithm,algorithmicx}
\usepackage{algpseudocode}
\usepackage[hyphens]{url} 
\usepackage[hidelinks]{hyperref}

\usepackage{marginnote}
\usepackage{empheq}
\usepackage{bm}
\usepackage{accents}
\usepackage{cite}
\usepackage{balance}
\usepackage{latexsym}
\usepackage{graphicx}
\usepackage{float}
\usepackage{colortbl}

\usepackage{multicol}
\usepackage{float}
\usepackage{cite}
\usepackage{cleveref}
\usepackage{aligned-overset}

\usepackage[normalem]{ulem} 

\usepackage[usenames,dvipsnames,svgnames,table]{xcolor}
\usepackage{pgfplots}
\usepackage{soul}

\usepackage{xtab, booktabs} 
\xentrystretch{-.25} 

\usepackage{tikz}
\usetikzlibrary{arrows.meta} 
\usetikzlibrary{fit, positioning}
\usetikzlibrary{calc}
\usetikzlibrary{decorations.pathmorphing} 

\tikzstyle{block} = [rectangle, minimum width=1cm, minimum height=1cm, text centered, draw=black]
\tikzstyle{tallblock} = [rectangle, minimum width=.5cm, minimum height=1cm, text centered, draw=black]
\tikzstyle{line} = [thick,-,>=stealth]
\tikzstyle{arrow} = [thick,->,>=stealth]
\tikzstyle{roundedblock} = [rectangle, minimum width=4cm, minimum height=2cm, text centered, draw=black, rounded corners=0.2cm]

\newcommand{\citep}{\cite}

\let\arg\relax
\DeclareMathOperator*{\arg}{\mathrm{arg}}

\newcommand{\bs}{\boldsymbol}
\newcommand{\mc}{\mathcal}
\newcommand{\bb}{\mathbb}
\newcommand{\R}{\bb R}
\newcommand{\N}{\mathbb{N}}

\DeclareMathAlphabet{\mathbbmsl}{U}{bbm}{m}{sl}

\DeclareMathOperator*{\argmin}{\operatorname{argmin}}

\newcommand{\diag}{\operatorname{diag}}
\newcommand{\blkdiag}{\operatorname{blkdiag}}
\newcommand{\col}{\operatorname{col}}
\newcommand{\row}{\operatorname{row}}

\newcommand{\Rmnum}[1]{\expandafter\@slowromancap\romannumeral #1@}
\newcommand{\eod}{\ensuremath{\hfill\Box}}
\newcommand{\qedd}{\ensuremath{\hfill \blacksquare}}
\newcommand{\bsigma}{\boldsymbol{\sigma}}

\newcommand{\smallsup}[1]{\text{\tiny \textnormal{#1}}} 

\newcommand{\bu}{\boldsymbol{u}}

\newcommand{\buol}{\bs{u}^{\smallsup{OL}}}
\newcommand{\uol}{{u}^{\smallsup{OL}}}
\newcommand{\bucl}{\bs{u}^{\smallsup{CL}}}
\newcommand{\ucl}{{u}^{\smallsup{CL}}}
\newcommand{\xcl}{{x}^{\smallsup{CL}}}
\newcommand{\xol}{{x}^{\smallsup{OL}}}
\newcommand{\Pol}{{P}^{\smallsup{OL}}}
\newcommand{\Pcl}{{P}^{\smallsup{CL}}}
\newcommand{\Kol}{{K}^{\smallsup{OL}}}
\newcommand{\Kcl}{{K}^{\smallsup{CL}}}
\newcommand{\Xol}{{\mathbb{X}}_f^{\smallsup{OL}}}
\newcommand{\Xcl}{{\mathbb{X}}_f^{\smallsup{CL}}}
\newcommand{\Aol}{\bar{A}^{\smallsup{OL}}}
\newcommand{\Acl}{\bar{A}^{\smallsup{CL}}}

\newcommand{\sigmacl}{{\sigma}^{\smallsup{CL}}}
\newcommand{\bsigmacl}{\boldsymbol{\sigma}^{\smallsup{CL}}}
\newcommand{\Plqr}{{P}^{\smallsup{LQR}}}
\newcommand{\Alqr}{\bar{A}^{\smallsup{LQR}}}
\newcommand{\Klqr}{{K}^{\smallsup{LQR}}}
\newcommand{\cS}{\mathcal{S}}

\usepackage{changepage}
\newcommand{\tildeb}[1]{\accentset{\sim}{#1}}

\newcommand{\xfh}{x^{\smallsup{FH}}}
\newcommand{\ufh}{u^{\smallsup{FH}}}
\newcommand{\bufh}{\bs{u}^{\smallsup{FH}}}

\newcommand{\buex}{\bs{u}^{\smallsup{EX}}}
\newcommand{\uex}{{u}^{\smallsup{EX}}}

\newcommand{\bush}{\bs{u}^{\smallsup{SH}}}
\newcommand{\ush}{{u}^{\smallsup{SH}}}

\makeatletter
\newcommand{\specialcell}[1]{\ifmeasuring@#1\else\omit$\displaystyle#1$\ignorespaces\fi}
\makeatother
\newtheorem{assumption}{Assumption}
\newtheorem{proposition}{Proposition}

\newtheorem{definition}{Definition}
\newtheorem{lemma}{Lemma}

\newtheorem{theorem}{Theorem}{\it}{}
\newtheorem{fact}{Fact}
\newtheorem{remark}{Remark}

\newcommand{\smallsum}{\mathop{\vcenter{\hbox{$\textstyle\sum$}}}\limits}
\newcommand{\tsum}{\textstyle\sum}

\newcommand{\VI}{\mathrm{VI}}

\newcommand{\comment}[1]{\textcolor{red}{[#1]}}

\title{Linear-Quadratic Dynamic Games as Receding-Horizon Variational Inequalities}
\author{Emilio Benenati, Sergio Grammatico
\thanks{Emilio Benenati is with the Division of Decisions and Control Systems (DCS), KTH, Stockholm, Sweden.
	S. Grammatico is with the Delft Center of Systems and Control (DCSC), TU Delft, the Netherlands. E-mail addresses: \texttt{benenati@kth.se, s.grammatico@tudelft.nl}. }
\thanks{This work was partially supported by the ERC under research project COSMOS (802348). }
}

\begin{document}

\maketitle
\begin{abstract}
	We consider dynamic games with linear dynamics and quadratic objective {functions.} We observe that the unconstrained open-loop Nash equilibrium coincides with a linear quadratic regulator in an augmented space, thus deriving an explicit expression of the cost-to-go. With such cost-to-go as a terminal cost, we show asymptotic stability for the receding-horizon solution of the finite-horizon, constrained game. {Furthermore, we show that the problem is equivalent to a non-symmetric variational inequality, which does not correspond to any Nash equilibrium problem. For unconstrained closed-loop Nash equilibria, we derive a receding-horizon controller that is equivalent to the infinite-horizon one and ensures asymptotic stability.}
\end{abstract}


\section{Introduction}
{We consider a {regulation} problem for a constrained, discrete-time linear-quadratic (LQ) dynamic game, which emerges when the agents have interest in reaching and maintaining a known attractor while optimizing an {individual} objective. Dynamic games (precisely, their continuous-time counterpart, \emph{differential} games), were first studied in the seminal paper \cite{starr_nonzero-sum_1969}.} They model a discrete-time dynamical system governed by multiple inputs, each controlled by a decision maker (or agent) with a self-interested objective. Applications include robotics \cite{lecleach_algames_2022, spica_real-time_2020}, robust control \cite{theodor_output-feedback_1996}, logistics planning \cite{hall_game-theoretic_2024} and {energy} markets \cite{hall_receding_2022}, to cite a few. 
{ An input strategy which no agent can improve via unilateral changes is called a Nash equilibrium (NE).} Interestingly, depending on the information structure of the problem, dynamic games may admit different types of NE. {In particular, if
each agent only observes the initial state and commits to a sequence of inputs which is
an optimal response to the input sequences of the other agents, then the strategy is an
open-loop Nash equilibrium (OL-NE); instead, if each agent is allowed to continuously
observe the state, then a closed-loop Nash equilibrium (CL-NE) is an optimal feedback
policy given the current state and the control policies of the other agents.} {It is well-known that a linear infinite-horizon CL-NE policy is linked to the solution of coupled AREs, see e.g. \cite[Proposition 6.3]{basar_dynamic_1999}. Recently, the authors of \cite{monti_feedback_2024} develop novel sufficient conditions for a similar OL-NE characterization.  In \cite{papavassilopoulos_linear-quadratic_1984}, the infinite-horizon continuous-time CL-NE problem was related to the problem of finding invariant linear subspaces: this observation has led to a significant development of the field \cite{engwerda_lq_2005} and solution algorithms based on a geometric approach \cite{engwerda_algorithms_2007}. Other algorithms, based on the  iterative solution of Lyapunov or Riccati equations have been presented in \cite{li_lyapunov_1995, freiling_global_1996, nortmann_nash_2024}. The recent works \cite{mylvaganam_constructive_2015, cappello_approximate_2020} study approximate solutions to differential games, with \cite{cappello_distributed_2022} providing a distributed computation approach. Crucially, none of these works consider the inclusion of constraints, and no established algorithmic method to compute infinite-horizon NE trajectories exists, with only \cite{nortmann_nash_2024} providing {a-posteriori,} local convergence guarantees for the discrete-time CL-NE setting.}  \par 
{Computational methods for the constrained, finite-horizon  case are instead available both for OL-NE \cite{lecleach_algames_2022} and CL-NE \cite{laine_computation_2023}. These have sparked the interest for receding-horizon controllers, which compute the NE for a finite-horizon game at each time step and apply the first input of the horizon as a control action. Such approach carries multiple advantages, as the recomputation allows the agents to react to unexpected disturbances, and the inclusion of constraints becomes tractable.} Indeed, receding-horizon solutions of dynamic games were successfully employed in autonomous racing \cite{spica_real-time_2020}, autonomous driving \cite{lecleach_algames_2022}, logistics planning \cite{hall_game-theoretic_2024} and electricity market clearance \cite{hall_receding_2022} applications. Generally, the stability properties of the closed-loop system are not analyzed, with the exception of \cite{hall_receding_2022} {and \cite{hall_stability_2024},} under the restrictions of considering a \emph{potential} game {and a \emph{stable} plant, respectively.} 
{The game-theoretic receding-horizon setup is also related to distributed, non-cooperative model predictive control (MPC). Early attempts at non-cooperating MPC formulations resulted in unstable closed-loop dynamics, see \cite[Sec. 4]{venkat_distributed_2008}, unless additional constraints enforce a cooperation between the agents \cite{camponogara_distributed_2002, magni_nonlinear_2009, dunbar_distributed_2007}. In our view, the introduction of these constraints do not fully capture the non-cooperative nature of the problem. Compared to the latter references, we do not introduce additional constraints; and compared to \cite{alessio_decentralized_2007, alessio_decentralized_2011, keviczky_decentralized_2006} we do not assume that the agents cooperate towards minimizing a common objective function. } \par
{Fundamentally, we show that the infinite-horizon unconstrained OL-NE can be interpreted as the input sequence generated by coupled linear quadratic regulator (LQR) problems in an augmented state space. Considering the problem in this augmented space is a genuinely novel perspective which yields to the first direct connection between the OL-NE and the LQR. Building on this novel insight, we derive a new closed-form expression for the value achieved by the OL-NE.} We include this expression as terminal cost for a receding horizon game-theoretic controller. With this terminal cost, the OL-NE trajectories coincide with the constrained infinite-horizon ones - a result which generalizes the infinite-horizon optimality property of single-agent MPC \cite{bei_hu_toward_2002}. {Furthermore, we show that the resulting controller is stabilizing: This is the first stability result for a game-theoretic predictive controller that enforces state and input constraints on a non-potential game.} From a computational perspective, we show that the finite-horizon problem can be cast as a Variational Inequality (VI) \cite{facchinei_generalized_2007}. The extensive literature on VIs allows one to design efficient, decentralized algorithms with convergence guarantees under loose assumptions. Interestingly, we find that the additive terminal cost jeopardizes the structure of the problem, as the resulting VI {(despite being solved by the truncation of an infinite-horizon NE) is not associated to the KKT conditions of any finite-horizon NE problem.} In the CL-NE case, the expression of the unconstrained, infinite-horizon cost-to-go is well-known in the literature. We formulate a receding-horizon control problem with such cost-to-go as terminal cost. We show that the receding-horizon solution coincides with the infinite-horizon one in the unconstrained case. Our technical contributions are summarized as follows:
\begin{itemize}
	\item {For the infinite-horizon, unconstrained OL-NE case, we propose a solution algorithm inspired by \cite{nortmann_nash_2024} based on the iterative solution of Stein equations {(Section \ref{sec:o-ne_inf_hor}).} We show that the OL-NE is equivalent to a set of LQRs in an augmented space and we derive a novel expression for its cost-to-go {(Section \ref{sec:cost_to_go}).}
	\item Leveraging the findings in Section \ref{sec:o-ne_inf_hor}, \ref{sec:cost_to_go}, we derive a finite-horizon constrained problem whose solutions coincide with the infinite-horizon one, thus generalizing the single-agent MPC infinite-horizon optimality property \cite{bei_hu_toward_2002}. We show that the system in closed loop with the receding-horizon solution of the considered finite-horizon problem is asymptotically stable and it can be cast as a VI (Section \ref{sec:rec_hor_olNE}).}
	\item By an appropriate choice of the terminal cost and prediction model, we derive a receding-horizon controller whose trajectories are equivalent to the ones of the infinite-horizon CL-NE in the unconstrained case (Section \ref{sec:rec_hor_clNE}). {Compared to the OL-NE case, the CL-NE solutions anticipate and account for the future reactions of the other players \cite{laine_computation_2023}, and this case admits an infinite-horizon stabilizing solution under less restrictive conditions.}
\end{itemize}
{In Section \ref{sec:numerics}, we illustrate two application examples via numerical simulations on distributed automatic generation control and vehicle platooning.}

\section{{Notation and Problem Statement}} 
\tabletail{\bottomrule}
\tablefirsthead{
  \toprule
  \textsc{Notation} & \textsc{Definition} \\ 
  \midrule
}
\tablehead{\bottomrule}
\renewcommand{\arraystretch}{1.06}

\begin{xtabular}{l | p{.62\columnwidth}}
	$\row(M_i)_{i\in \mc I}$  & row stack: $\begin{bmatrix} M_1,\dots, M_N\end{bmatrix}$ \\ 
	\hline 
	$\col(M_i)_{i\in \mc I}$  & column stack: $\begin{bmatrix} M_1^\top,\dots, M_N^\top\end{bmatrix}^\top$ \\
	\hline
	$\blkdiag(M_i)_{i\in \mc I}$ & diagonal stack of $(M_i)_{i\in\mc I}$. \\
		\hline
	$\|x\|_M$ & weighted Euclidean norm: $\sqrt{x^{\top}Mx}$\\
		\hline
	$\mc{S}^n_{T}$ & set of $T$-long sequences in $\R^n$  \\
		\hline
	$v[t]$ & $t$-th element of $v\in\mc{S}^n_{T}$\\
		\hline
	$v[t|p]$ & for a function $p\mapsto v(p) \in\mc{S}^n_{T}$, denotes the $t$-th element of $v(p)$ \\
	\hline
	$f(x|y)$ & for each value of a parameter $y$, denotes the function \quad\quad $f(\cdot|y):x\mapsto f(x|y)$  \\
			\hline
	$\delta_i^K$ & vector in $\R^K$ with only non-zero element $1$ at index $i$ \\
	\hline
	$I_k$ & identity matrix of size $k$\\
		\hline
	$N$ & number of agents \\ 
		\hline
	$n$ & dimension of state variable \\
		\hline
	$m$ & dimension of input variable \\
		\hline
	$\mc I$ & set of all agents: $\{1,...,N\}$ \\	
	\hline
	$\mc I_{-i}$  & set of all agents except $i$: $\mc I \setminus \{i\}$\\	
	\hline
	$T$ & length of the control horizon  \\ 	\hline
	$\mc T$ & control horizon: $\{0,..., T-1\}$ \\	\hline
	$\mc T^+$  & shifted horizon: $\{1,..., T\}$ \\ 	\hline
	$x$ & state variable \\	\hline
	$u_i$ & input sequence of agent $i$ \\	\hline
	$\bu$ & collective input sequence: $(u_i)_{i\in\mc I}$ \\ 	\hline
	$\bu_{-i}$ & input sequence for all agents except $i$: $(u_i)_{i\in\mc I_{-i}}$ \\ 	\hline
	$\phi(x_0, \bu)$ & state sequence of system \eqref{eq:dynamics} with initial state $x_0$ and input $\bu$\\	\hline
	$Q_i$ & state weight matrix for agent $i$ \\	\hline
	$R_i$ & input weight matrix for agent $i$ \\	\hline
	$C_i$ & Square root of $Q_i$: $Q_i=C_i^{\top}C_i$ \\	\hline
	$\ell_i$ & stage cost for agent $i$, see \eqref{eq:stage_cost} \\	\hline
	$S_i$ & $:=B_iR_i^{-1}B_i^{\top}$ \\	\hline
	$\mathbb{X}$ & set of admissible states \\	\hline
	$\mathbb{U}_i$ &  maps to the set of admissible inputs for agent $i$, see \eqref{eq:input_constraints} \\	\hline
	$\mc{U}_{i,T}$ & maps to the set of $T$-long admissible input sequences for agent $i$, see \eqref{eq:feas_set_ith} \\	\hline
	$\bs{\mc{U}}_{T}$ & maps to the set of $T$-long admissible collective input sequences, see \eqref{eq:feas_set_collective} \\	\hline
	$\VI(\mc F, \mc C)$ & variational inequality \cite{facchinei_finite-dimensional_2007}: for $\mc C$ closed, convex set and operator $\mc F$, find $v^*\in \mc C$ such that  \\
		& $\inf_{v \in \mc C}\langle \mc F(v^*), v - v^*\rangle \geq 0$.\\
		\bottomrule
	\end{xtabular} \\
\par
We consider the problem of regulating to the origin the state of the dynamical system 
\begin{equation}\label{eq:dynamics}
	x[t+1] = Ax[t] + \smallsum_{i\in\mc I} B_i u_i[t],
\end{equation}
 subject to  state and coupling input constraints 
\begin{subequations}
\begin{align}
x[t]&\in\mathbb{X}\subseteq \R^{n}, & \forall{t}\in\mc T; \label{eq:state_constraints}  \\
u_i[t]&\in\mathbb{U}_i(\bu_{-i}[t]) \subseteq \R^{m} & \forall{t}\in\mc T, ~i\in\mc I. \label{eq:input_constraints}
\end{align} 
\end{subequations}
Each input is determined by a self-interested agent. Without loss of generality, we consider inputs with equal dimension $m$ for each agent.
We denote the state sequence of the system resulting by \eqref{eq:dynamics} with initial state $x_0$ and collective input sequence $\bu = (u_i)_{i\in\mc I}$ as $\phi(x_0,\bu)$. {According to the notation, we denote the state evolution at time $t$ as $\phi[t|x_0, \bu]$.}
{We consider control problems over a horizon $T$, possibly infinite, with quadratic stage costs
\begin{align}\label{eq:stage_cost}
	\begin{split}
		\forall i\in\mc I: ~ & \ell_i(x,u_i)= \tfrac{1}{2}  \|x\|^2_{Q_i} + \tfrac{1}{2} \|u_i\|^2_{R_i}. 
	\end{split}
\end{align}}
Define for each $i$ the feasible input sequences
 \begin{align}\label{eq:feas_set_ith}\begin{split}
 	\mc{U}_{i,T}(x_0, \bu_{-i}):=\{&u_i\in\cS^{m}_T| \\
 	&u_i[t]\in\mathbb{U}_i(\bu_{-i}[t])~\forall t\in\mc T;\\
 	&	\phi[t|x_0, \bu]\in\mathbb{X} 
 	~\forall t\in\mc T^+\} \end{split}
 \end{align}
 and the collective input sequences 
 \begin{align}\label{eq:feas_set_collective}
 	\bs{\mc{U}}_T(x_0):= \{\bs{u}\in\cS_T^{Nm}| u_i \in \mc{U}_{i,T}(x_0, \bu_{-i}) ~\forall i\in\mc I \}.
 \end{align}
 	Finally, we assume that the origin is strictly feasible and that the state and input weights are positive semi-definite and definite, respectively.
\begin{assumption}\leavevmode \label{as:basic_assm} 
		\begin{enumerate}[label=(\roman*), ref={\theassumption(\textit{\roman*}})]
		\item \label{as:basic_assm:constraints} $0\in\mathrm{int}(\mathbb{X})$;  $\forall i\in \mc{I}, 0\in\mathrm{int}(\mathbb{U}_i(0))$.
		\item \label{as:basic_assm:cost} $Q_i = C_i^{\top}C_i \succcurlyeq 0, ~ R_i = R_i^{\top} \succ 0 \quad \forall i \in \mc I.	$
	\end{enumerate}
 \end{assumption}
 
\section{{Open-loop Nash trajectories}} \label{sec:o-ne}
Depending on the information structure of the problem assumed for the agents, dynamic games can have different solutions concepts \cite[Ch. 6]{haurie_games_2012}. {In this section, we consider the infinite-horizon open-loop Nash equilibrium (OL-NE) trajectories, where}
each agent assumes that the opponents only observe the initial state and subsequently commit to the sequence of inputs computed using such observations. Define the objective
\begin{equation}\label{eq:inf_hor_objective}
	\forall i:~J^{\infty}_i(u_i|x_0, \bs{u}_{-i}):= \smallsum_{t=0}^{\infty} \ell_i(\phi[t|x_0, \bu], u_i[t]).
\end{equation}
The OL-NE trajectory is defined as follows:
\begin{definition}\label{def:ol_ne}\cite[Def. 6.2]{haurie_games_2012}:
	Let $x_0\in\mathbb{X}$. The sequences $\bs u\in\bs{\mc{U}}_\infty(x_0)$ are an open-loop Nash trajectory at $x_0$ if, for all $i\in\mc I$,
	\begin{equation}\label{eq:one_def}
		J^{\infty}_i(u_i|x_0, \bs u_{-i}) \leq \inf_{v_i\in\mc{U}_{i,\infty}(x_0, \bs u_{-i})} J^{\infty}_i(v_i|x_0, \bs u_{-i}).
	\end{equation}
\end{definition}
\subsection{The unconstrained infinite-horizon case}\label{sec:o-ne_inf_hor} 
Let us first consider the unconstrained OL-NE problem, which is related to the Riccati equations  
{
\begin{subequations}\label{eq:riccati_open_loop}
	\begin{empheq}[left={\forall i\in\mc I:\empheqlbrace\,}]{align}
		\Pol_i &= Q_i + A^{\top}\Pol_i \Aol \label{eq:riccati_open_loop:P} \\[1ex]
		\Kol_i &= -R_i^{-1}B_i^{\top} \Pol_i \Aol, \label{eq:riccati_open_loop:K}
	\end{empheq}
\end{subequations}}
where
\begin{equation} \label{eq:closed_loop_A_explicit}
\Aol := A + \smallsum_{i\in\mc I} B_i \Kol_i.
\end{equation}
{By leveraging \cite[Theorem 4.10]{monti_feedback_2024}, a solution to \eqref{eq:riccati_open_loop} defines an OL-NE trajectory, as we report next: }
  \begin{assumption}\cite[Assumptions 4.6, 4.7]{monti_feedback_2024}\label{as:ol_ne_primitives}
  	    \renewcommand\theassumption{\arabic{assumption}}
  	\begin{enumerate}[label=(\roman*), ref={\theassumption(\textit{\roman*})}]
 	\item\label{as:ol_ne_primitives:A_inv} The matrix $A$ is invertible. 
 	\item\label{as:ol_ne_primitives:stab_det} For all $i\in\mc I$, the pairs $(A, B_i)$ and $(A, C_i)$ introduced in \eqref{eq:dynamics} and Assumption \ref{as:basic_assm} are respectively stabilizable and detectable.
 	\end{enumerate}
 \end{assumption}
\begin{assumption}\cite[Assumption 4.9]{monti_feedback_2024} \label{as:structure_H}
	The matrix
	\begin{equation}\label{eq:def_H}
		H:= \begin{bmatrix}
			A + \tsum_{j \in \mc I} \large(S_j A^{-\top} Q_j\large) & \row(-S_j A^{-\top})_{j\in\mc I}\\
			\col(-A^{-\top}Q_j)_{j\in\mc I} & I_N \otimes A^{-\top}
		\end{bmatrix}
	\end{equation}
	possesses exactly $n$ eigenvalues  with modulus smaller than $1$. Moreover, an $n$-dimensional stable invariant subspace of $H$ is complementary to
	\begin{equation*}
		\mathrm{Im}\left(\begin{bmatrix}
			\bs{0}_{n\times Nn} \\ 
			I_{Nn} 
		\end{bmatrix} \right).
	\end{equation*}
\end{assumption}
\begin{proposition}\label{prop:one_characterization}
	Let Assumptions \ref{as:basic_assm}, \ref{as:ol_ne_primitives}, \ref{as:structure_H} hold true and let $(\Kol_i, \Pol_i)_{i\in\mc I}$ satisfy \eqref{eq:riccati_open_loop}. {Let $\Aol$ be defined as in \eqref{eq:closed_loop_A_explicit}.} For any $x_0\in\R^{n}$, let $\buol(x_0)\in\mc S_{\infty}^{Nm}$, $\xol(x_0)\in\mc S_{\infty}^{n}$ be defined as {
	\begin{subequations} \label{eq:def_ol_sequence}
		\begin{align}
		 \label{eq:def_xol} 
				\forall  t\in\N_0:~&\xol[t|x_0] := (\Aol)^t x_0 \\
				\forall i \in \mc I, t\in\N_0:~ &\uol_i[t|x_0] := \Kol_i \xol[t|x_0].  \label{eq:def_uol}
		\end{align}
	\end{subequations} }	
 Then, {$\buol(x_0)$} is an OL-NE trajectory at the initial state $x_0$ {and $\Aol$ is Schur.}
\end{proposition}
\begin{proof}
	See Appendix \ref{app:o-ne}.
\end{proof}
{
 If we disregard the dependence of $\Aol$ on $(\Pol_j)_{j\in\mc I}$, \eqref{eq:riccati_open_loop:P} is a set of $N$ Stein equations, for which off-the-shelf solvers exist \cite{van_huffel_high-performance_2004}. One can then iteratively find $(\Pol_i)_{i\in\mc I} $ that solves Equation \eqref{eq:riccati_open_loop:P}  with $\Aol$ fixed, and then update $\Aol$ according to
	\begin{equation} \label{eq:closed_loop_A}
		\Aol=  (I + \tsum_{j\in\mc I} S_j\Pol_j)^{-1} A.
	\end{equation}
	The equality in \eqref{eq:closed_loop_A} is proven in Fact \ref{fct:reform_closed_loop_mat} (Appendix \ref{app:o-ne}) when $A$ is not singular. This approach is formalized in Algorithm \ref{alg:open_loop_sylvester}.}
 \begin{algorithm}\caption{OL-NE solution via Stein recursion}\label{alg:open_loop_sylvester} 
	\begin{algorithmic}[1]
  {		\State \textbf{Initialization:} $(K_i^{(0)})_{i\in\mc I}$ such that $\bar{A}^{(0)}=A + \tsum_{i\in\mc I} B_iK_i^{(0)}$ is Schur;
 		\For{$k\in\N$}: 
 		\For{$i\in\mc I$}:
 		\State Solve $P_i^{(k+1)} = Q_i + A^{\top}P_i^{(k+1)}\bar{A}^{(k)};$
 		\EndFor
 		\State $\bar{A}^{(k+1)} \gets (I + \tsum_{j\in\mc I} S_jP_j^{(k+1)})^{-1} A$
 		\For{$i\in\mc I$}:
\State  $ K_i^{(k+1)} \gets -R_i^{-1}B_i^{\top} P_i^{(k+1)} \bar{A}^{(k+1)}.$
 		\EndFor
 		\EndFor}
 	\end{algorithmic}
 \end{algorithm} 
  
 \subsection{The OL-NE as a linear quadratic regulator}\label{sec:cost_to_go}
In this section, we show that the matrices $(\Kol_i)_{i\in\mc I}$ in \eqref{eq:def_uol} that characterize the unconstrained OL-NE are related to $N$ LQR controllers defined in a higher-dimensional space. This result offers a novel perspective on OL-NE and it is instrumental for our main results in Section \ref{sec:rec_hor_olNE}. {For each agent $i$, let us denote with $\Plqr_i, \Klqr_i$ the solution to the ARE that solves the standard LQR problem for the LTI system $(A,B_i)$, namely:}
\begin{subequations}\label{eq:lqr}
	\begin{align}
		\Plqr_i &= Q_i + A^{\top} \Plqr_i \Alqr_i; \label{eq:lqr:P}\\
		\Klqr_i &= -R_i^{-1}B_i^{\top} \Plqr_i \Alqr_i; \label{eq:lqr:K}\\
		\Alqr_i &:= A + B_i\Klqr_i. \label{eq:A_cl_lqr}
	\end{align}
\end{subequations}
In \cite{monti_feedback_2024} the authors note that, for each agent $i$, the OL-NE is the optimal control for the system $(A,B_i)$ perturbed by the actions of the other agents. We observe that, along the trajectory defined by the {control laws} $(\Kol_i)_{i\in\mc I}$, such perturbation is fully determined by the initial state and by the dynamics of the autonomous system $\Aol$. Remarkably, the problem can then be cast as an augmented regulator problem by considering a system with $2n$ states that incorporates the dynamics of the perturbation.

\begin{lemma}[Nash Equilibrium as augmented LQR solution]\label{le:one_is_lqr} Let {$(\Pol_i, \Kol_i)_{i\in\mc I}$} solve \eqref{eq:riccati_open_loop} and let $\Aol$ as in \eqref{eq:closed_loop_A_explicit}. Let Assumptions \ref{as:basic_assm:cost}, \ref{as:ol_ne_primitives}, \ref{as:structure_H} hold true. For all $i\in\mc I$, define
	\begin{subequations}
		\begin{align}
				\label{eq:auxiliary_lti_system}
			&	\widehat{A}_i:=\begin{bmatrix}
				A  & \sum_{j\neq i} B_j \Kol_j \\
				0 & \Aol
			\end{bmatrix},  \quad \widehat{B}_i := \begin{bmatrix}
			B_i \\ 
			0
			\end{bmatrix} \\
			&{\widehat{Q}_i := \begin{bmatrix}Q_i & 0 \\ 0 & 0 \end{bmatrix},} \quad \widehat{R}_i := R_i. \label{eq:auxiliary_lti_system:objective}
	\end{align}
	\end{subequations}
	Then, the ARE
	\begin{subequations}\label{eq:riccati_expanded}
			\begin{align}
			\widehat{P}_i &= \widehat{Q}_i + \widehat{A}_i^{\top}\widehat{P}_i(\widehat{A}_i + \widehat{B}_i \widehat{K}_i) \label{eq:riccati_expanded:P} \\
			\widehat{K}_i&= -(\widehat{R}_i + \widehat{B}_i^{\top}\widehat{P}_i\widehat{B}_i)^{-1}\widehat{B}_i^{\top}\widehat{P}_i\widehat{A}_i \label{eq:riccati_expanded:K}
		\end{align}
	\end{subequations}
	admits a unique positive semidefinite solution
	\begin{equation}\label{eq:solution_lifted_sys}
		\widehat{P}_i = \begin{bmatrix}
			\Plqr_i & \tildeb{P}_i \\
			\tildeb{P}_i^{\top} & *
		\end{bmatrix}; \quad \widehat{K}_i = \begin{bmatrix}
		\Klqr_i & \tildeb{K}_i
		\end{bmatrix}
	\end{equation}
	where $\tildeb{P}_i = \Pol_i - \Plqr_i$ and $\tildeb{K}_i = \Kol_i - \Klqr_i$.
	Furthermore, for all $i\in\mc I$, $\widehat{A}_i + \widehat{B}_i\widehat{K}_i$ is Schur.\eod
\end{lemma}
\begin{proof}
	See Appendix \ref{app:cost_to_go}.
\end{proof}
{Let us consider the lifted system $(\widehat{A}_i, \widehat{B}_i)$ in \eqref{eq:auxiliary_lti_system} {for some $i\in\mc I$} controlled by $\widehat{K}_i$ in \eqref{eq:solution_lifted_sys} with initial state $\col(x_0, y_0)$. 
{Denote by $\col(x^*, y^*)$ and $u_i^*$ the resulting state and input sequences, that is,}
\begin{subequations}
	\begin{align}
			u^*_i[t] &= \Klqr_i x^*[t] + \tildeb{K}_i y^*[t]; \\
			x^*[t+1] &= Ax^*[t] + B_iu^*_i[t]+  \tsum_{j\neq i} B_j {\Kol_j}y^*[t]; \label{eq:lifted_dyn_x}\\
			y^*[t+1] &= \Aol y^*[t]. \label{eq:lifted_dyn_y}
	\end{align}
\end{subequations}
Note that \eqref{eq:lifted_dyn_y} implies $y^*[t] = (\Aol)^t y_0$. Furthermore, from \eqref{eq:def_ol_sequence}, $\uol_j[t|y_0] = \Kol_j(\Aol)^t y_0$, for all $j\in\mc I$.  Thus, substituting the latter in \eqref{eq:lifted_dyn_x},
\begin{align*} 
	\begin{split} 
	x^*[t+1] =& Ax^*[t] + B_iu_i^*[t] + \smallsum_{j\neq i} B_j \uol_j[t|y_0].
\end{split}
\end{align*}
{In other words, $x^*$ is the sequence of states of the (non-lifted) dynamics in \eqref{eq:dynamics} controlled by $u_i^*$ and $\buol_{-i}(y_0)$. We write this compactly as }
\begin{equation} \label{eq:x_star_is_seq_of_nonlifted_sys_compact}
	x^*[t] = \phi[t|x_0, u_i^*, \buol_{-i}(y_0)].
\end{equation} 
Consider the function
\begin{align}\label{eq:def_V}
	V_i(x_0, y_0):= \frac{1}{2} \left\| \begin{bmatrix}
		x_0 \\ y_0
	\end{bmatrix}\right\|_{\widehat{P}_i}^2
\end{align}
where $\widehat{P}_i$ solves \eqref{eq:riccati_expanded}. {By applying a known result in the LQR literature \cite[Thm. 21.1]{hespanha_linear_2018},
\reversemarginpar

	\begin{align*} 
	\begin{split}
		\frac{1}{2} \left\| \begin{bmatrix}
			x_0 \\ y_0
		\end{bmatrix}\right\|_{\widehat{P}_i}^2&{=} \frac{1}{2} \smallsum_{t=0}^{\infty} 
	\left\|\begin{bmatrix}x^*[t] \\ y^*[t] \end{bmatrix}\right\|^2_{\widehat{Q}_i} + \left\|u_i^*[t]\right\|^2_{\widehat{R}_i}.
	\end{split} 
	\end{align*}
	By substituting \eqref{eq:auxiliary_lti_system:objective} in the latter to eliminate $\widehat{Q}_i, \widehat{R}_i$,}  we find:
	\begin{align}\label{eq:V_optimal_xy}
	\begin{split}
		V_i(&x_0, y_0)= \tfrac{1}{2}\smallsum_{t=0}^{\infty} \|x^*[t]\|_{Q_i}^2 + \|u_i^*[t]\|_{R_i}^2 \\
		 \overset{\eqref{eq:x_star_is_seq_of_nonlifted_sys_compact}, \eqref{eq:stage_cost}}&{=}\tfrac{1}{2}\smallsum_{t=0}^{\infty} \ell_i(\phi[t|x_0, u^*_{i}, \buol_{-i}(y_0)], u_i^*[t]) \\
		\overset{\eqref{eq:inf_hor_objective}}&{=} J^{\infty}_i(u_i^*|x_0, \buol_{-i}(y_0)) \\
		&\leq J^{\infty}_i(u_i|x_0, \buol_{-i}(y_0)) \quad \forall  u_i\in\mc{S}_{\infty}^{m}
	\end{split}
\end{align} 
where the latter inequality follows from the optimality of $u_i^*$. In particular, for $x_0=y_0$, we have
\begin{equation}\label{eq:V_optimal_xx}
	V_i(x_0,x_0) \leq J^{\infty}_i(u_i|x_0, \buol_{-i}(x_0)) ~~ \forall u_i\in\mc{S}_{\infty}^{m}.
\end{equation}}
{Note that, from Definition \ref{def:ol_ne}, for all $u_i\in\mc{S}_{\infty}^{m}$ and $\buol$ defined as in \eqref{eq:def_ol_sequence},
\begin{equation}
	J^{\infty}_i(\uol_i(x_0)|x_0, \buol_{-i}(x_0)) \leq J^{\infty}_i(u_i|x_0, \buol_{-i}(x_0)).
\end{equation}
 Thus, comparing the latter with \eqref{eq:V_optimal_xx}, we conclude that $V_i(x_0,x_0)$ is the infinite-horizon cost-to-go of the OL-NE trajectory $\buol$ from state $x_0$:}
\begin{equation}\label{eq:V_is_gne_cost_to_go}
	V_i(x_0,x_0) = J^{\infty}_i(\uol_i(x_0)|x_0, \buol_{-i}(x_0)).
\end{equation}
Finally, we observe from the Bellman optimality principle applied to the system $(\widehat{A}_i, \widehat{B}_i)$ that
\begin{multline}\label{eq:V_value_function}
		V_i(x_0, y_0) = \min_{u_i\in\R^{m}} \tfrac{1}{2}\large(\|x_0\|^2_{Q_i} + \|u_i\|^2_{R_i} \large) +  \\
		V_i(Ax_0+B_iu_i +\tsum_{j\neq i} B_j \Kol_j y_0, \Aol y_0)
\end{multline} 
and the latter optimization problem has solution $\Klqr_ix_0 + \tildeb{K}_iy_0$.
\begin{figure}
	\centering\resizebox{\columnwidth}{!}{\begin{tikzpicture}[>=latex, node distance=1.1cm, auto]

    \node (ui) at (0,0) {$u_i[k]$};
    \node [draw, block, right=.8cmof ui] (Bi) {${B_i}$};
    \node [draw, circle, inner sep=1pt, right=1cm of Bi] (sum) {$+$}; 
    \node [draw, tallblock, right=.6cmof sum] (delay) {$\frac{1}{z}$};
    \node [fill=black, circle, right=.6cmof delay, inner sep=1pt] (dot) {};
    \node [above=2pt of dot.west] {$x[k]$};
    \node [right=.6cmof dot] (exit) {};
    \node [draw, block, below=.2cmof delay] (A) {${A}$};

    \node [draw, block, above=2.5cm of Bi.west] (Knoti) {$(B_j\Kol_j)_{j\neq i}$};
    \node [fill=black, circle, left=0.3cmof Knoti, inner sep=1pt] (dot2) {};
    \node [draw, block, left=0.3cmof dot2] (Abar) {$\Aol$};
    \node [draw, tallblock, below=.2cm of Abar] (delay2) {$\frac{1}{z}$};
    \node [
        draw,
        dashed,
        color=red,
        line width=1pt, 
        fit=(Knoti) (Abar) (delay2),
        inner sep=0.3cm,
    ] (group) {};

    \node [
        draw,
        dashed,
        color=blue,
        line width=1pt, 
        fit=(delay) (A) (sum) (dot),
        inner sep=0.3cm,
    ] (group) {};

    \draw[->] (ui.east) --  (Bi.west);
    \draw[->] (Bi.east) -- (sum.west);
    \draw[->] (sum.east) -- (delay.west);
    \draw[-] (delay.east) -- node[above] {} (dot.west);
    \draw[->] (dot.east) -- (exit.west);
    \draw[->] (dot.south) |- (A.east);
    \draw[->] (A.west) -| (sum.south);
    \draw[->] (Knoti.east) -| node[above] {$\buol_{-i}[k]$}  (sum.north);
    \draw[->] (dot2.east) -- (Knoti.west);
    \draw[-] (Abar.east) -- (dot2.west);
    \draw[->] (dot2.south) |- (delay2.east);
    \draw[->] (delay2.west) -- ++(-0.45,0) |- (Abar.west);

\end{tikzpicture}}
	\caption{{Block diagram of the augmented system in \eqref{eq:auxiliary_lti_system} for agent $i$: the red dashed line highlights an uncontrollable mode which generates $\buol_{-i}$. Following \eqref{eq:V_optimal_xx}, $\uol_i$ is the optimal control for this system if both the subsystems highlighted in red and in blue have the same initial state.}}
\end{figure}
 
 \subsection{Receding horizon Open-Loop Nash equilibria} \label{sec:rec_hor_olNE}
 {In this section we construct a finite-horizon problem whose solution is a truncation of the constrained infinite-horizon OL-NE. This result, shown in Theorem \ref{thm:fin_hor_ne_is_inf_hor_ne}, generalizes the infinite-horizon optimality property of single-agent MPC, which is known to require a careful tuning of the terminal cost function \cite{bei_hu_toward_2002}. Intuitively, the finite-horizon problem in \eqref{eq:def_u_star} is defined by including the infinite-horizon cost-to-go of the OL-NE $V_i$ in \eqref{eq:def_V} as a terminal cost: however, as $V_i$ is defined in an augmented space, special care is needed. We show in Theorem \ref{thm:stability_closed_loop} that, with this design choice, the receding-horizon control law obtained by applying the first input of the solution to the finite-horizon problem makes the origin asymptotically stable. Finally, in Proposition \ref{prop:one_as_vi}, we cast the finite-horizon problem as a Variational Inequality (VI). Consider the problem, parametrized in $x_0\in\mathbb{X}$, of finding $\bufh(x_0)$ such that}
 \begin{subequations}
\begin{align}\label{eq:def_u_star}
		&\forall~ i\in\mc I:~ \ufh_{i}(x_0) \in \argmin_{u_i\in \mc U_{T,i}(x_0, \bufh_{-i})} J_i(u_i|x_0, \bufh), \\
	& J_i(u_i|x_0, \bufh):= \tsum_{t\in\mc T} \left( \ell_i(\phi[t|x_0, u_i,\bufh_{-i}], u_i[t])\right)\nonumber\\ 
	&\quad+ V_i\big(\phi[T|x_0, u_i, \bufh_{-i}], \phi[T|x_0, \bufh]\big), \label{eq:cost_rec_hor_ol} 
\end{align} 
\end{subequations}
{where $V_i$ is as in \eqref{eq:def_V}. Note that, for each agent $i$, the optimization problem in \eqref{eq:def_u_star} is parametric in the decision variables of \emph{all} agents $\bufh$. Interestingly, this structure differs from the one of a static NE problem, where the parametrization is on the decision variables of the \emph{remaining} agents, see e.g. \cite[Eq. 1]{facchinei_generalized_2010}.  }
Let $\Xol$ be a constraints-admissible forward invariant set for the dynamics $x[t+1] = \Aol x[t]$. {Techniques for computing $\Xol$ can be found in \cite{darup_computation_2014} and references therein.} Define the set
\begin{align}\label{eq:reachable_set}
	\begin{split}
			\mc X:=\big\{x_0\in\mathbb{X}~|~ &\exists \bufh~ \text{that solves} ~\eqref{eq:def_u_star}; \\
					&\phi[T|x_0, \bufh]\in\Xol\big\}.
	\end{split}
\end{align}
{We show next that the infinite-horizon constrained OL-NE trajectory can be recovered by the solutions to \eqref{eq:def_u_star}.}
\begin{theorem}\label{thm:fin_hor_ne_is_inf_hor_ne}
	Let Assumptions \ref{as:basic_assm}, \ref{as:ol_ne_primitives}, \ref{as:structure_H} hold true. Let $(\Pol_i, \Kol_i)_{i\in\mc I}$ solve \eqref{eq:riccati_open_loop} and let $\buol(x)\in \cS^{Nm}_{\infty}$ be the unconstrained OL-NE sequence for any $x\in\mathbb{X}$, as defined in \eqref{eq:def_ol_sequence}. Let $\bufh\in \cS^{Nm}_{T}$ solve \eqref{eq:def_u_star} for some $x_0\in\mc X$, with associated state sequence $\xfh := \phi\left(x_0, \bufh\right)$, and  define the \emph{extended input sequence} $\buex \in\mc  S^{Nm}_{\infty}$ as
	\begin{equation}\label{eq:def_u_star_inf_ol}
		\forall i:~\uex_i[t]:=\begin{cases}
			\ufh_i[t] & \text{if} ~t < T\\
			\uol_{i}[t-T|\xfh[T]] & \text{if} ~t \geq T.
		\end{cases}
	\end{equation}
	 Then, $\buex$
	is an infinite-horizon OL-NE trajectory for the system in \eqref{eq:dynamics} with state and input constraint sets $\mathbb{X}, (\mathbb{U}_i)_{i\in\mc I}$, respectively, and initial state $x_0$.
\end{theorem} 
\begin{proof}
	See Appendix \ref{app:rec-hor-one}.
\end{proof}
In light of Proposition \ref{thm:fin_hor_ne_is_inf_hor_ne}, by solving \eqref{eq:def_u_star} at subsequent time instants, one expects the agents not to deviate from the previously found solution. This is because they will recover shifted truncations of the same infinite-horizon OL-NE trajectory. {Indeed, in Lemma \ref{le:shifted_sequence_is_solution} we show that a trajectory solving the problem in \eqref{eq:def_u_star} when shifted by one time step still solves the problem in \eqref{eq:def_u_star} for the subsequent state.} {This is crucial, because the control action remains the same between subsequent computations of the solution, keeping the evaluated objective constant--except for the first stage cost, which does not appear in the summation. The cumulative objective of the agents decreases then at each time step and it can be used as a Lyapunov function to show the stability of the origin (modulo some technicalities, due to the fact that $Q_i$ is only positive semidefinite for all $i$). Let us formalize this next.  }
\begin{lemma}\label{le:shifted_sequence_is_solution}
	Let Assumptions \ref{as:basic_assm}, \ref{as:ol_ne_primitives}, \ref{as:structure_H} hold. Let $x_0\in\mc X$, with $\mc X$ defined in \eqref{eq:reachable_set}. Define the \emph{shifted input sequence} $\bush\in\mc S_T^{Nm}$ as follows:
	\begin{equation}\label{eq:shifted_sol}
	\forall i\in\mc I:~	\ush_{i}[t] = \begin{cases}
			\ufh_{i}[t+1] & \text{if} ~ t < T-1 \\
			\Kol_i \xfh [T] & \text{if} ~ t = T-1
		\end{cases}
	\end{equation}
	where $\bufh$ solves \eqref{eq:def_u_star} with initial state $x_0$, {$\xfh := \phi\left(x_0, \bufh\right)$} {and $(\Pol_i, \Kol_i)_{i\in\mc I}$ solve \eqref{eq:riccati_open_loop}.}
	Then, $\bush$ is a solution for the problem in \eqref{eq:def_u_star} with initial state $\xfh[1]$. 
\end{lemma}
\begin{proof}
	See Appendix \ref{app:rec-hor-one}. 
\end{proof}

In view of Lemma \ref{le:shifted_sequence_is_solution} and given that the problem in \eqref{eq:def_u_star} might admit multiple solutions, we assume that the shifted solution in \eqref{eq:shifted_sol} is actually employed when the agents solve subsequent instances of the problem in \eqref{eq:def_u_star}. Assumption \ref{as:select_shifted_seq} is practically reasonable as, when implementing a solution algorithm for \eqref{eq:def_u_star}, one can warm-start the algorithm to the shifted sequence $\bush$ defined in \eqref{eq:shifted_sol}. 
\begin{assumption}\label{as:select_shifted_seq}
	For any $x_0\in\mathbb{X}$, if the problem in \eqref{eq:def_u_star} at the initial state $x_0$  admits a solution $\bufh$ and if the shifted sequence $\bush$ defined in \eqref{eq:shifted_sol} is a solution of \eqref{eq:def_u_star} with initial state $\phi[1|x_0, \bufh]$, then  $\bush$ is selected by all agents when solving  \eqref{eq:def_u_star} with initial state $\phi[1|x_0, \bufh]$.
\end{assumption}
We are now ready to conclude on asymptotic stability of the system controlled in receding horizon:
\begin{theorem} \label{thm:stability_closed_loop}
	Let Assumptions \ref{as:basic_assm}--\ref{as:ol_ne_primitives}, \ref{as:structure_H}--\ref{as:select_shifted_seq} hold true. Consider the system in \eqref{eq:dynamics} with feedback control $x\mapsto\bufh[0|x]$, where $\bufh(x)$ solves \eqref{eq:def_u_star} for the initial state $x$. The origin is asymptotically stable for the closed-loop system with region of attraction $\mc{X}$, defined in \eqref{eq:reachable_set}.
\end{theorem}
\begin{proof}
	See Appendix \ref{app:rec-hor-one}. 
\end{proof}

\subsection{The open-loop Nash equilibrium as a Variational Inequality}\label{sec:vi_one}
Theorem \ref{thm:fin_hor_ne_is_inf_hor_ne} bridges the infinite-horizon constrained OL-NE trajectory with the solution to a finite-horizon equilibrium problem, whose solutions can be computed algorithmically. In fact, we recast the problem in \eqref{eq:def_u_star} as a Variational Inequality (VI), for which a plethora of efficient solution algorithms exist under some standard monotonicity and convexity assumptions \cite{facchinei_finite-dimensional_2007}.
\begin{proposition} \label{prop:one_as_vi}
	Assume $\bs{\mc{U}}_T(x_0)$ non-empty, closed and convex for all $x_0\in\mathbb{X}$. For some $T\in\N$, define for all $i\in\mc I$: 
	\begin{align}\begin{split} \label{eq:matrices_VI}
		\Gamma_{i} &:= \begin{bmatrix} 
			B_i & 0 & \dots & 0 \\
			AB_i & B_i & \dots & 0\\
			\vdots &  & \ddots & \\
			A^{T-1} B_i & A^{T-2} B_i  & \dots & B_i
		\end{bmatrix},\\
		\Theta&:= \col(A^k)_{k\in\mc T^+};\\
		 \bar{R}_{i}&:= I_T \otimes R_i \\
		\bar{Q}_{i}&:= \begin{bmatrix}
			I_{T-1} \otimes Q_i & 0 \\
			0 & \Pol_i
		\end{bmatrix} \end{split}
	\end{align}
	and define $F(\cdot|x_0):\R^{TNm} \to \R^{TNm}$, parametric in $x_0\in\R^n$, as
	\begin{align}\label{eq:linear_condition_fin_hor}
		\begin{split}
			&F(\bu|x_0) := \blkdiag(\bar{R}_{i})_{i\in\mc I}\bu +\\
			&  \begin{bmatrix}
				\Gamma_{1}^{\top}\bar{Q}_{1} \\ 
				\vdots \\
				\Gamma_{N}^{\top}\bar{Q}_{N} 
			\end{bmatrix} \begin{bmatrix}\Gamma_{1} &  \dots & \Gamma_{N}  \end{bmatrix}\bu  +\begin{bmatrix}
				\Gamma_{1}^{\top}\bar{Q}_{1} \\ 
				\vdots \\
				\Gamma_{N}^{\top}\bar{Q}_{N}
			\end{bmatrix} \Theta x_0.
		\end{split} 
	\end{align}
	Then, any solution of $\VI(F(\cdot|x_0), \bs{\mc{U}}_T(x_0) )$ is a solution to \eqref{eq:def_u_star}.
\end{proposition}
\begin{proof}
	See Appendix \ref{app:rec-hor-one}.
\end{proof}
 It can be shown via the the generalized Gerschgorin disk theorem \cite{feingold_block_1962} that the VI in Proposition \ref{prop:one_as_vi} is strongly monotone \cite[Def. 2.3.1]{facchinei_finite-dimensional_2007} if, for all $i$, $R_i=r_iI$, with $r_i>0$ large enough. A strongly monotone VI admits a unique solution \cite[2.3.3]{facchinei_finite-dimensional_2007}, thus this design choice makes Assumption \ref{as:select_shifted_seq} redundant. {The proposed control algorithm is summarized in Algorithm \ref{alg:ol-ne_control}. We refer to \cite{baghbadorani_douglas-rachford_2025} for a runtime benchmark and implementation of different solution algorithms available to perform step 3.}
\begin{remark}\label{re:finite_hor_NE} {It is well-known that a static convex game can be cast as a VI defined via the stacked partial gradients of the agents cost functions \cite{facchinei_generalized_2007}.
	We note, however, that the matrix multiplying $\bu$ in \eqref{eq:linear_condition_fin_hor} has non-symmetric diagonal blocks, since $(\Pol_i)_{i\in\mc I}$ are in general non-symmetric matrices. This implies that there do not exist $N$ cost functions (one for each block) such that $F$ corresponds to their stacked gradients. Therefore, there is no static game whose NE corresponds to the solution of \eqref{eq:def_u_star}.} {This confirms the observation made in Section \ref{sec:rec_hor_olNE} that the structure of \eqref{eq:def_u_star} differs from the one of a static NE problem.}
\end{remark} 
{\begin{algorithm}\caption{Receding horizon OL-NE control} \label{alg:ol-ne_control}
	\begin{algorithmic}[1]
		{\For{$t\in\N$} 
			\State Measure $x[t]$ 
			\State Find $\bufh$ that solves $\mathrm{VI}(F(\cdot|x[t]), \bs{\mc U}_T(x[t]))$
			\For{$i\in\mc I$}
			\State Apply the input $\ufh_i[0]$
			\EndFor
			\EndFor
		}
	\end{algorithmic}
\end{algorithm}}
  
\section{{Closed-Loop Nash equilibria}} \label{sec:rec_hor_clNE}
We now turn our attention to the closed-loop Nash equilibrium (CL-NE) solution concept, where each agent assumes that the opponents can observe the state at each time step and recompute their input sequence accordingly. The game is defined over the feedback control functions $\sigma_i: \mathbb{N} \times \mathbb{R}^{n}\to \mathbb{R}^{m}$. {Denote $\bsigma=(\sigma_i)_{i\in\mc I}$. We overload the notation for the state sequence of the system in \eqref{eq:dynamics} controlled by the feedback law $\bsigma$ as $\phi(x_0, \bsigma)$. Furthermore, we denote as $u_i(x_0, \bsigma)$ the sequence of inputs for agent $i\in\mc I$ resulting from $\bsigma$: that is, for all $t\in\mc{T}$,
\begin{align*}
	u_i[t|x_0, \bsigma] &= \sigma_i(t, \phi[t|x_0, \bsigma]), ~~~~ \forall  i\in\mc I;\\
	\phi[t+1|x_0, \bsigma] &= A\phi[t|x_0, \bsigma] + \tsum_{i\in\mc I} B_i u_i[t|x_0, \bsigma].
\end{align*} 
We also overload the notation for the objective in \eqref{eq:inf_hor_objective} as }
\begin{align} \label{eq:cost_feedback}
	\begin{split}
		\forall& i\in\mc I: ~ 	J^{\infty}_i(\sigma_i|x_0, \bsigma_{-i})= \\
		& \qquad\qquad\smallsum_{t=0}^{\infty} \ell_i(\phi[t| x_0, \bsigma], u_i[t|x_0, \bsigma]).
	\end{split}
\end{align}
We consider the unconstrained case with $\mathbb{X}=\R^n$, $\mathbb{U}_i=\R^m$, for all $i\in\mc I$.
\begin{definition}\cite[Def. 6.3]{haurie_games_2012}:
	The feedback strategies $\bsigma^*:\R^n\to\R^{Nm} $ are a CL-NE if for all $x_0\in\R^n$, for all $i\in\mc I$,
	\begin{align}
		\begin{split}
			J^{\infty}_i(\sigma^*_i|x_0, \bsigma^*_{-i}) &\leq \inf_{\sigma_i:\R^n\to\R^{m}} J^{\infty}_i(\sigma_i|x_0, \bsigma^*_{-i}). 
		\end{split}
	\end{align}
\end{definition}
\begin{figure}
	\centering\resizebox{.8\columnwidth}{!}{\begin{tikzpicture}[>=latex, node distance=1cm, auto]

    \node (ui) at (0,0) {};
    \node [draw, block, right=of ui] (Bi) {${B_i}$};
    \node [draw, circle, inner sep=1pt, right=0.6cm of Bi] (sum) {$+$}; 
    \node [draw, tallblock, right=1.2cmof sum] (delay) {$\frac{1}{z}$};
    \node [fill=black, circle, right=3cmof sum, inner sep=1pt] (dot) {};

    \node [draw=none, right=.6cmof dot] (yi) {};
    \node [draw, block, below=.3cm of delay] (A) {${A}$};
    \node [draw, block, below=.3cm of A] (Ki) {$\Kcl_i$};

    \node [draw, block, above=.5cmof sum] (Bnoti) {$(B_j)_{j\neq i}$};
    \node [draw,block, right=of Bnoti, , xshift=2mm] (Knoti) {$\Kcl_{-i}$};

    \draw[->] (ui.east) -- node[above] {${u}_{i}[k]$} (Bi.west);
    \draw[->] (Bi.east) -- (sum.west);
    \draw[->] (sum.east) -- (delay.west);
    \draw[-] (delay.east) -- node[above] {$x[k]$} (dot.west);
    \draw[->] (dot.south) |- (A.east);
    \draw[->] (dot.south) |- (Ki.east);
    \draw[-] (Ki.west) -| (ui.east);

    \draw[->] (dot.north) |- (Knoti.east);
    \draw[->] (A.west) -| (sum.south);
    \draw[->] (dot.east) -- (yi.west);
    \draw[->] (Knoti.west) -- node[above] {$\boldsymbol{u}_{-i}[k]$} (Bnoti.east);
    \draw[->] (Bnoti.south) -| (sum.north);
    
\end{tikzpicture}}
	\reversemarginpar}{\caption{Block diagram of the CL-NE control problem for agent $i$. The linear feedback $\Kcl_i$ is optimal for the depicted system. }
\end{figure}
{The unconstrained CL-NE problem admits a linear feedback solution. We report a known characterization and stability result for the CL-NE in terms of coupled AREs, which we then employ to design a stabilizing receding-horizon controller in Section \ref{sec:rec_hor_clNE}.}
\begin{assumption}\label{as:reachability_controllability_clNE}
	$(A, \row(B_i)_{i\in\mc I})$ is  stabilizable and $(A, \sum_{i\in\mc I} Q_i)$ is detectable.
\end{assumption}
\begin{lemma} \cite[Cor. 3.3]{monti_feedback_2024} \label{le:cl_NE_characterization} Let Assumption \ref{as:basic_assm:cost}, \ref{as:reachability_controllability_clNE} hold true. Let  $(\Pcl_i, \Kcl_i)_{i\in\mc I}$ solve 
		\begin{subequations}\label{eq:riccati_CL} 
			    \begin{empheq}[left={\forall i\in\mc I:\empheqlbrace\,}]{align} 
			\Pcl_i &= Q_i + (\Acl_{-i})^{\top} \Pcl_i \Acl \label{eq:riccati_CL:P}  \\
			\Kcl_i &= - R_i^{-1} B_i^{\top} \Pcl_i \Acl \label{eq:riccati_CL:K} 
		    \end{empheq}
		\end{subequations}
	with $\Pcl_i = (\Pcl_i)^{\top} \succcurlyeq 0$, 
	\begin{align}\label{eq:A_closed_loop_cl}
		\begin{split}
			\Acl &:= A + \tsum_{j\in\mc I} B_j \Kcl_j\\
			\Acl_{-i} &:= A + \tsum_{j\neq i} B_j \Kcl_j.
		\end{split}		
	\end{align} Then, the linear feedback control $\bsigma=(\Kcl_i)_{i\in\mc I}$ is a CL-NE, the resulting closed-loop dynamics is asymptotically stable and $J_i(\sigma_i|x_0, \bsigma_{-i})=\frac{1}{2}\|x_0\|_{\Pcl_i}^2$ for all $i\in\mc I$ and $x_0\in\R^{n}$.
\end{lemma}
\begin{proof}
	It follows directly via algebraic calculations from \cite[Cor. 3.3]{monti_feedback_2024}.
\end{proof}
If one ignores the dependence of $\Acl_{-i}$ on $\Kcl_i$ (which emerges via \eqref{eq:riccati_CL:K}), \eqref{eq:riccati_CL:P} is a standard Riccati equation that solves the LQR problem with state evolution matrix $\Acl_{-i}$ and thus one can expect its solution to be symmetric. For this reason, \eqref{eq:riccati_CL} is sometimes referred to in the literature as a symmetric coupled ARE, as opposed to \eqref{eq:riccati_open_loop}, whose solutions are in general not symmetric. As proposed in \cite{nortmann_nash_2024}, the CL-NE can either be computed by iteratively fixing $(\Kcl_{j})_{j\neq i}$ for each $i$ and solving \eqref{eq:riccati_CL:P} with a Riccati equation solver, or by rewriting \eqref{eq:riccati_CL:P} as
\begin{align}
	\begin{split}
	\Pcl_i &= Q_i + (\Acl)^{\top}\Pcl_i\Acl - (\Kcl_i)^{\top}B_i^{\top} \Pcl_i \Acl  \nonumber \\
		\overset{\eqref{eq:riccati_CL:K}}&{=} \underbrace{Q_i + (\Pcl_i\Acl)^{\top}S_i \Pcl_i \Acl}_{\tildeb{Q}_i} + (\Acl)^{\top}\Pcl_i\Acl  \nonumber\\
	\end{split}
\end{align}
and by solving the latter via a Lyapunov equation solver, considering $\tildeb{Q}_i$ fixed at each iteration.
\subsection{Receding horizon closed-loop Nash equilibria}
In this section, we study the stability of the origin for the system \eqref{eq:dynamics} in closed-loop with the receding-horizon solution of a finite-horizon CL-NE problem. As a CL-NE is a feedback law valid on the whole state space, one needs not recomputing it at each iteration and thus the concept of a receding-horizon CL-NE is counterintuitive. However, {one must consider that} the state-of-the-art solution algorithm for finite-horizon CL-NE problems \cite{laine_computation_2023}, which handles state and input constraints, only computes {a single trajectory} resulting from a CL-NE given an initial state. By receding-horizon CL-NE, we then mean computing at each time step a trajectory resulting from a finite-horizon CL-NE {policy,} and applying the first input of the sequence. For stability purposes, a natural choice for the terminal cost is $\|\cdot\|_{\Pcl_i}^2$, where $(\Kcl_i, \Pcl_i)_{i\in\mc I} $ solve \eqref{eq:riccati_CL}. The finite-horizon CL-NE problem can be written as a NE problem with nested equilibrium constraints parametrized in the initial state  \cite[Theorem 2.2]{laine_computation_2023}:
\begin{subequations}\label{eq:nested_gne}
		\begin{alignat}{3}
			\intertext{$\forall i:~ \sigma_{T,i}^*(x[0])\in$}
			&\arg_{u_i[0]}&\min_{u_i\in\mc{S}_{T}^m}&\tfrac{1}{2}\|x[T]\|^2_{\Pcl_i} + \smallsum_{t=0}^{T-1} \ell_i(x[t], u_i[t])  \label{eq:nested_gne:cost} \\
			 & &\text{s.t.} ~  & x[t+1] = Ax[t] + B_i u_i[t] \nonumber \\
			 & & & \qquad ~\quad+ \smallsum_{j\in\mc I_{-i}} B_j \sigma^*_{T-t,j}(x[t]), ~~ \forall t\in\mc T.  \label{eq:nested_gne:dyn}
		\end{alignat} 
\end{subequations}
{In the latter, the notation $\arg_{u_i[0]}\min_{u_i\in\mc{S}_{T}^m}$ denotes that the minimum is taken over $u_i$, while only the first element $u_i[0]$ is returned.} The equilibrium constraints emerge in \eqref{eq:nested_gne:dyn}, as $\bsigma^*_{T-t}$ is defined by an instance of the problem in \eqref{eq:nested_gne} {over the shortened horizon $T-t$.} 
{The equilibrium constraints cannot be expressed explicitly as a constraint set, and thus \eqref{eq:nested_gne} cannot be directly written as a VI.  We then consider a surrogate problem, obtained by substituting $\sigma_{T-t,j}^*$ with the linear feedback $\Kcl_{j}$ for all $t\in\mc T,j\neq i$. The resulting problem only presents standard affine coupling constraints in the game decision variables, which allows for a VI reformulation.}
\begin{subequations}\label{eq:surrogate_gne}
	\begin{alignat}{3}
		\intertext{$\forall i:~	u^*_i(x[0]) \in$}
		&\arg&\min_{u_i\in\mc{S}_{T}^m}&\tfrac{1}{2}\|x[T]\|^2_{\Pcl_i} + \smallsum_{t=0}^{T-1} \ell_i(x[t], u_i[t])  \label{eq:surrogate_gne:cost}  \\
		&             &              \text{s.t.}~&x[1]= Ax[0]+ B_i u_i[0] + \tsum_{j\neq i} B_j u_j[0]; \label{eq:surrogate_gne:dyn_0} \\
		&             &			   				&x[t+1]= \Acl_{-i}x[t]+ B_i u_i[t], ~~ \forall t\neq 0. \label{eq:surrogate_gne:dyn}
	\end{alignat} 
\end{subequations}

We remark that, given $x[0]$, \eqref{eq:surrogate_gne} is a static NE problem (without equilibrium constraints) {\reversemarginpar and it thus admits a VI reformulation via standard results \cite{facchinei_generalized_2010}.} {Intuitively, this substitution modifies the prediction model of each agent: instead of expecting the remaining agents to apply a finite-horizon CL-NE with shrinking horizon, it is predicted that they will apply the infinite-horizon CL-NE. Next, we show that the choice of prediction model and terminal cost ensures that $(\Kcl_i x_0)_{i\in\mc I}$ is a solution to \eqref{eq:surrogate_gne} from any initial state $x_0$.}
\begin{theorem} \label{thm:convergence_cl_NE}
	Let Assumptions  \ref{as:basic_assm:cost}, \ref{as:reachability_controllability_clNE} hold. For all $T\in\N$, $x_0\in\R^n$, {$\bucl$ is a solution to \eqref{eq:surrogate_gne}} for the initial state $x_0$, defined as
 {\begin{align} \label{eq:surrogate_solution} \begin{split}
	\xcl[t] &:= (\Acl)^t x_0 \qquad \forall t \in \mc T;\\
 	\ucl_i[t] &:= \Kcl_i \xcl[t] \qquad \forall t \in \mc T,~\forall i\in\mc I,
 \end{split}
 \end{align}}
{where $(\Kcl_i)_{i\in\mc I}$ solve \eqref{eq:riccati_CL} and $\Acl$ is in \eqref{eq:A_closed_loop_cl}.}
\end{theorem}
\begin{proof}
	See Appendix \ref{app:rec_hor_clNE}.
\end{proof}
{The stability of the origin under the receding-horizon controller follows then from the one of $(\Kcl_i)_{i\in\mc I}$. Note that the results of Theorem \ref{thm:convergence_cl_NE} trivially hold also when input and state constraints are included, if the initial state is in a constraints-admissible forward invariant set for the autonomous system $x[t+1] = \Acl x[t]$, thus still ensuring local asymptotic stability of the origin. {This is because, from Theorem \eqref{thm:convergence_cl_NE}, the input $u_i^*$ defined in \eqref{eq:surrogate_solution} is the  minimizer for the $i$-th unconstrained problem in \eqref{eq:surrogate_gne} when the other agents apply the input $\bs{u}_{-i}^*$. Since $u_i^*$ is constraint-admissible, it is also a minimizer of the constrained problem: note that the introduction of the constraints does not modify the model of the dynamics assumed by agent $i$.  \normalmarginpar The same considerations do not hold for the case in \eqref{eq:nested_gne}, because the equilibrium constraints in \eqref{eq:nested_gne:dyn} are affected by the introduction of state and input constraints, thus $\bs{u}^*$ might not  satisfy the equation in \eqref{eq:nested_gne:dyn} even when it is constraint-admissible for the state and input constraints. Intuitively speaking, the feedback $(K_i)_{i\in\mc I}$ might not be a CL-NE even when state-and-input constraint admissible, as one agent could ``take advantage'' of the opponents' constraints by driving the state outside of the constraint-admissible region.}}

\section{Application examples$^1$} \label{sec:numerics}
\subsection{Vehicle platooning}
\let\thefootnote\relax\footnote{\noindent$^1$Code available at  \protect\url{github.com/bemilio/Receding-Horizon-GNE}}
We consider the vehicle platooning scenario in \cite{shi_distributed_2017}. The leading vehicle, indexed by $1$, aims at reaching a reference speed $v^{\mathrm{ref}}$, while the remaining agents $i\in\{2,...,N\}$ aim at matching the speed of the preceding vehicle, while maintaining a desired distance $d_i$, plus an additional speed-dependent term $h_iv_i$, where $v_i$ is the speed of agent $i$ and $h_i$ is a design parameter. For all $i\neq 1$, the local state is 
\begin{equation}
	x_i = \begin{bmatrix}
		p_{i-1} - p_i - d_i - h_iv_i \\
		v_{i-1} - v_i
	\end{bmatrix},
\end{equation} 
where $p_i$ denotes the position of agent $i$ with respect to the one of agent $1$. As the position of agent $1$ with respect to itself is $0$, we define $x_1 =\begin{bmatrix} 0,
	v^{\mathrm{ref}}-v_1\end{bmatrix}^\top$.
The dynamics is that of a single integrator and $N-1$ double integrators sampled with rate $\tau_{\text{s}}=0.1s$. With algebraic calculations, we obtain
\begin{align} \label{eq:platooning_dyn}
	\begin{split}
		A &= \blkdiag\left(\begin{bmatrix}
			0 & 0 \\
			0 & 1
		\end{bmatrix}, I_{N-1} \otimes\begin{bmatrix}
			1 & \tau_{\text{s}} \\
			0 & 1
		\end{bmatrix} \right); \\
		B_1 &=  \delta_{2}^N \otimes \begin{bmatrix}
			\tau_{\text{s}}^2/2 \\ 
			\tau_{\text{s}}
		\end{bmatrix} - \delta_1^N \otimes \begin{bmatrix}
			0 \\ 
			\tau_{\text{s}}
		\end{bmatrix};\\
		B_N&= -\delta_N^N \otimes \begin{bmatrix}
			h_i\tau_{\text{s}} + \tau_{\text{s}}^2/2 \\ 
			\tau_{\text{s}}
		\end{bmatrix};\\
		\forall i&\in\{2,...,N-1\}: \\
		B_i& =  \delta_{i+1}^N \otimes \begin{bmatrix}
			\tau_{\text{s}}^2/2 \\ 
			\tau_{\text{s}}
		\end{bmatrix} - \delta_i^N \otimes \begin{bmatrix}
			h_i\tau_{\text{s}} + \tau_{\text{s}}^2/2 \\ 
			\tau_{\text{s}}
		\end{bmatrix},
	\end{split}
\end{align}
where $ \delta^n_i\in\R^n$ is a vector with only non-zero element $1$ at index $i$. We impose the following safety distance, speed and input constraints: 
\begin{align} \label{eq:speed_input_constraints}
	\begin{split}
		p_{i-1}  &\geq d^{\text{min}}_i +  p_i; \\
		v_i^{\text{min}} &\leq v_i \leq v_i^{\text{max}};\\
		u_i^{\text{min}} &\leq u_i \leq u_i^{\text{max}}.
	\end{split}
\end{align}
As the system in \eqref{eq:platooning_dyn} does not satisfy the stabilizability assumption \ref{as:ol_ne_primitives:stab_det}, we apply to each agent $i$ a pre-stabilizing local controller 
\begin{equation*}
	K_i^{\text{stab}} = (\delta_i^N)^{\top} \otimes [-1, -1].
\end{equation*}
We then apply the OL-NE receding horizon control with state and input weights $Q_i = I$, $R_i =1$ and horizon $T=10$.  {The VI defined in Proposition \ref{prop:one_as_vi} is solved via the forward-backward splitting method \cite[\S 12.4.2]{facchinei_computation_2011}, see also \cite[Algorithm 1]{belgioioso_distributed_2022} for an implementation in the context of NE problems. This method employs dualization of the constraints which couple the decision variables of each agents (i.e. state constraints and coupling input constraints), while the local input constraints are handled via a projection step.} A sample trajectory is shown in Figure \ref{fig:platooning_pos_vel}, where we observe that the vehicles achieve the desired equilibrium state while satisfying all the constraints. We compute a suitable set $\Xol$ by inscribing a level set of a quadratic Lyapunov function of the autonomous system with dynamics $\Aol$ in the polyhedron defined by \eqref{eq:speed_input_constraints}. As shown in Figure \ref{fig:platooning_pos_vel}(c), the state enters the set $\mc X$ defined in \eqref{eq:reachable_set}  after $t=11.7 s$: {At this instant, all conditions of Theorem \ref{thm:stability_closed_loop} are satisfied and convergence to the origin is guaranteed.} Furthermore,  we verify numerically that, for each subsequent time step $\tau$, the input sequences $\bush$ computed at time $\tau$ as in \eqref{eq:shifted_sol} are a solution for the game at time step $\tau+1$, which is to be expected due to Lemma \ref{le:shifted_sequence_is_solution}. {We report that we are currently unable to compute an accurate a-priori estimate of the region of attraction, and this test shows that it is larger than the set $\mc X$. {Additionally, we test the robustness of the method to errors in the terminal cost function. We perturb the matrices $(\Pol_i)_{i\in\mc I}$ with an additive error matrix, generated by sampling a normal distribution. We perform $100$ state realizations for each tested value of the variance. In Figure \ref{fig:box_plot_noise} we plot the difference between the resulting state trajectory and the nominal state trajectory $\hat{x}$. We observe that the problem is robust to small perturbations of the terminal cost function, as the state sequence deviates by approximately 2\% when the variance is 1\% of the maximum element of $(\Pol_i)_{i\in\mc I}$. The error increases significantly for increasing error values. We observe that the system remains asymptotically stable in all the tests.}
\begin{figure}
	\centering
	\includegraphics[width=\columnwidth]{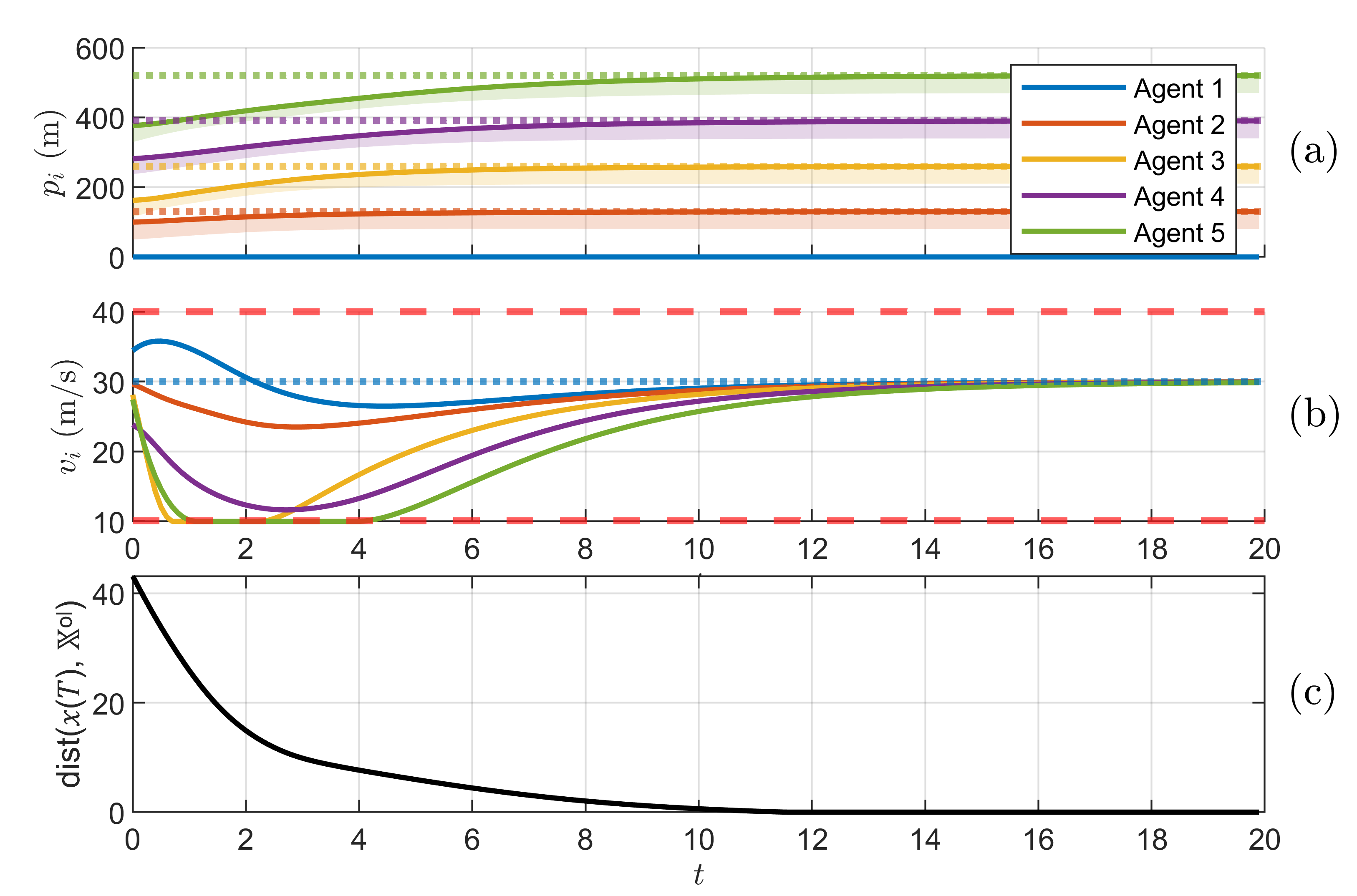}
	\caption{{Vehicle platooning test case. (a) Position with respect to the leading agent. The shaded areas in the first plot represent the distance constraints for each agent. The dotted lines represent the desired values at steady state. (b) Velocity. The dotted lines represent the desired values at steady state. The dashed red lines represent the constraints. (c) Distance of the terminal state to $\Xol$.  }  }
	\label{fig:platooning_pos_vel}
\end{figure}
\begin{figure}
    \centering
    \includegraphics[width=\columnwidth]{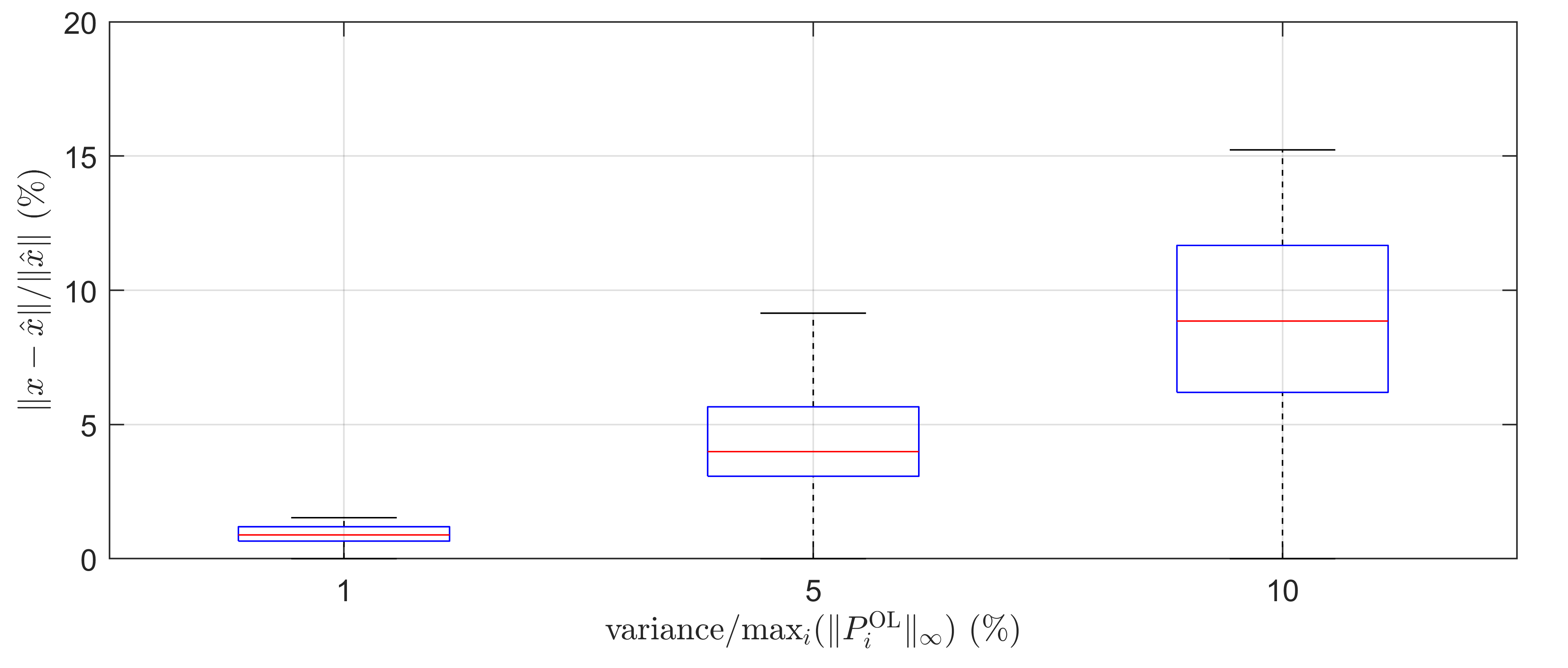}
    \caption{{Deviation between the nominal state sequence $\hat{x}$ and the state sequence $x$ resulting from the receding horizon solution of (26) when the matrices $(\Pol_i)_{i\in\mc I}$ are perturbed by an additive error sampled from a normal distribution.} }
    \label{fig:box_plot_noise}
\end{figure}
\subsection{Distributed control of interconnected generators}
\label{sec:power_system}
We test the inclusion of the proposed terminal costs on the automatic generation control problem for the power system application considered in \cite{venkat_distributed_2008}. The power system under consideration is composed of $4$ generators interconnected via tie-lines arranged in a line graph. {The models for the dynamics of the generators and of the tie lines linearized around a steady-state reference are the ones in \cite[Eq. 17]{venkat_distributed_2008}, with the model parameters specified in \cite[\S A.1]{venkat_distributed_2012}}. This application example is chosen as it is observed in \cite{venkat_distributed_2008} that the distributed MPC control architecture (equivalent to the receding-horizon OL-NE without terminal cost in this paper) leads to the system being unstable and thus it is a challenging distributed control problem. The model considered has $3$ states for each generator (namely, the angular velocity of the rotating element, the mechanical power applied to the rotating element and the position of the steam valve) and one for each tie line (namely, the power flow). Each agent has control authority over the reference point of their respective governor. The control objective for each agent is to regulate the deviation from the reference angular speed of the generator rotating part and power flow at the tie-line they are connected to, with the exception of agent 1 that does not control the tie line. In numbers, we have
\begin{align}
	\begin{split}
	R_i &= 1 \quad\forall i\in\{1,2,3,4\};\\
	Q_1 &= \blkdiag(5, \bs{0}_{14\times14})\\
	\forall i&\in\{2,3,4\}:\\ 
	Q_i& = \begin{bmatrix}
		\delta_i^4 (\delta_i^4)^{\top} \otimes \diag(5,0,0) & \bs{0}_{12\times 3}\\
		\bs{0}_{3\times 12}& 5 \delta_{i+1}^3 (\delta_{i+1}^3)^{\top} 
	\end{bmatrix}.
\end{split}
\end{align}
  We observe that {the system does not satisfy Assumption \ref{as:structure_H},} thus we could not test the performance of the OL-NE receding horizon controller. Conversely and as expected from Lemma \ref{le:cl_NE_characterization}, {we find a stabilizing infinite-horizon CL-NE using the iterations in \cite[Eq. 16]{nortmann_nash_2024}.}
 We test the receding-horizon CL-NE controller which solves the problem in \eqref{eq:surrogate_gne} from a randomly generated initial state, and we compare its stabilizing property with respect to the controller which implements \eqref{eq:surrogate_gne} without a terminal cost in Figure \ref{fig:cl_ne_percentage_stable}. We estimate a constrained admissible forward invariant set of the form 
 \begin{align}\label{eq:x_f_cl_power_system}
 	\begin{split}
 	\Xcl = \{x\in\mathbb{X}|\|x\|_P^2<r\}
 	\end{split}
 \end{align}
 where $P$ defines a quadratic Lyapunov function for the autonomous system $\Acl$ defined as in \eqref{eq:A_closed_loop_cl} and $r$ is determined numerically such that the controller $(\Kcl_i)_{i\in\mc I}$ is feasible. We observe that the system is asymptotically stable when the initial state is in $\Xcl$.  In general, the inclusion of the terminal cost is beneficial for the asymptotic convergence of the closed-loop system.
 \begin{figure}
 	\centering
 	\includegraphics[width=\columnwidth]{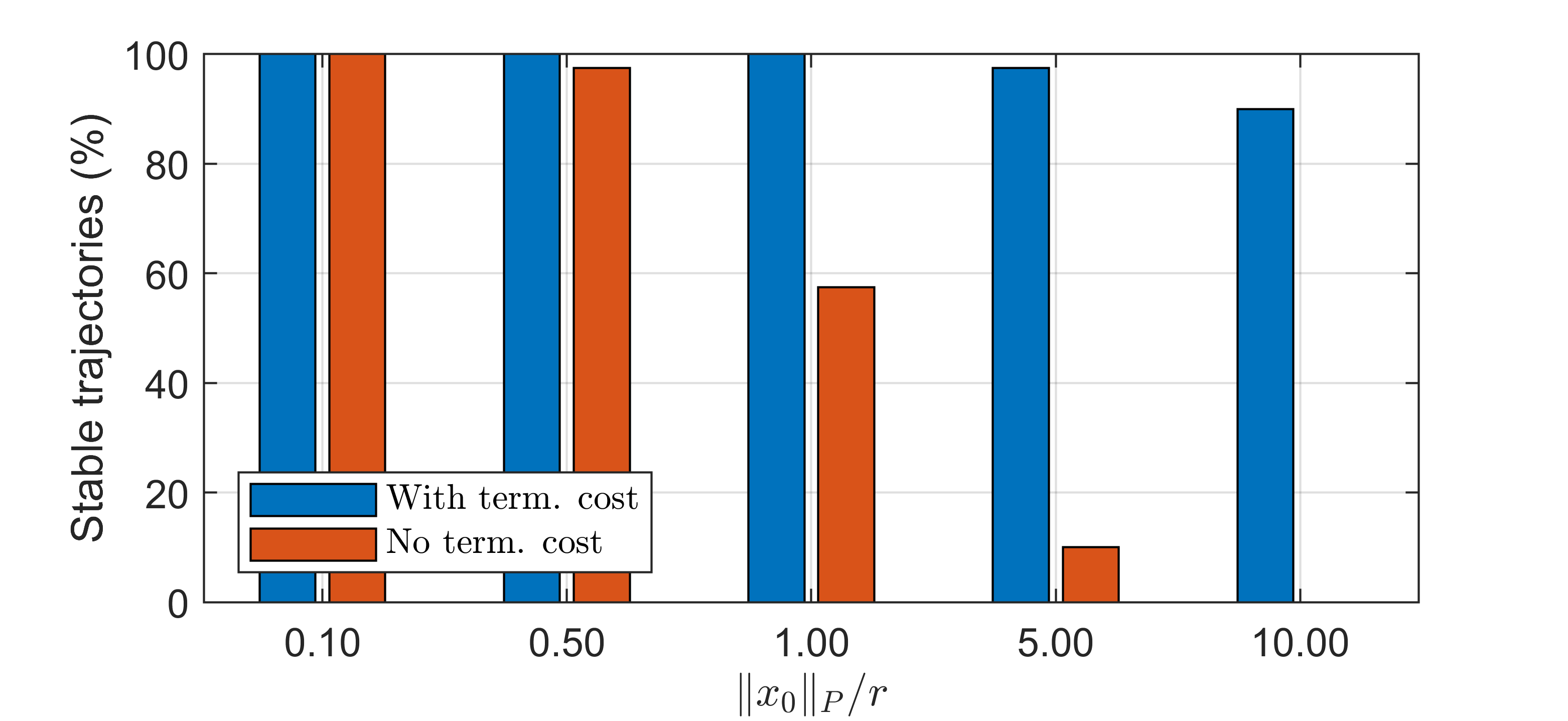}
 	\caption{Percentage of asymptotically stable trajectories for the 4-zone power systems in \cite{venkat_distributed_2008} controlled by the receding-horizon CL-NE for different initial states $x_0$. $P$ and $r$ define a constraint-admissible forward invariant set as in \eqref{eq:x_f_cl_power_system}. }
 	\label{fig:cl_ne_percentage_stable}
 \end{figure}

\section{Conclusion}
{The constrained infinite-horizon open-loop Nash equilibrium can be computed by solving a finite-horizon control problem, if the infinite-horizon cost-to-go {over an augmented state-space} is used as terminal cost function. The receding-horizon solution is asymptotically stable and the problem can be cast as a variational inequality.} Instead, the closed-loop Nash equilibrium case requires solving a game with nested equilibrium constraints. With an appropriate relaxation, we find a finite-horizon problem whose solution coincides with the infinite-horizon unconstrained closed-loop Nash equilibrium. {Open problems left for future work include developing algorithms with convergence guarantees for computing infinite-horizon unconstrained Nash equilibria, studying the constrained receding-horizon closed-loop Nash equilibrium case, {and determining a characterization of the monotonicity of the finite-horizon variational inequalities with respect to the problem data.} }

\appendix
\subsection{Additional results and proofs to Section \ref{sec:o-ne_inf_hor}} \label{app:o-ne}
\paragraph*{Proof of Proposition \ref{prop:one_characterization}}
{The proof is based on showing that \eqref{eq:riccati_open_loop} implies \cite[Eq. 4.32]{monti_feedback_2024} and then applying \cite[Thm. 4.10]{monti_feedback_2024}.} We left-multiply \eqref{eq:riccati_open_loop:P} by $S_iA^{-\top}$ and sum over $\mc I$ to obtain:
	\begin{align}\label{eq:equivalence_freiling_sassano:1}
		\begin{split}
			\smallsum_{j\in\mc I} S_j A^{-\top}(Q_j-\Pol_j) &=  \smallsum_{j\in\mc I} -S_j\Pol_j\Aol\\ \overset{\eqref{eq:riccati_open_loop:K}}&{=} \smallsum_{j\in\mc I} B_j\Kol_j.
		\end{split}
	\end{align}
	Thus,
	\begin{align}\label{eq:equivalence_freiling_sassano:4}
		\begin{split}
			\Aol 
		 \overset{\eqref{eq:closed_loop_A_explicit},\eqref{eq:equivalence_freiling_sassano:1}}&{=}  A + \tsum_{j\in\mc I} S_j A^{-\top}(Q_j-\Pol_j).
		\end{split}
	\end{align}
	By substituting \eqref{eq:equivalence_freiling_sassano:4} in \eqref{eq:riccati_open_loop:P} we find
	\begin{subequations}
	\begin{align}
		&A^{-\top} (Q_i - \Pol_i) = \nonumber\\
 		 &\qquad- \Pol_i(A + {\tsum_{j\in\mc I}}S_jA^{-\top}(Q_j-\Pol_j))\\
		&\Kol_i = -R_i^{-1}B_i^{\top}\Pol_i(A + \tsum_{j\in\mc I} S_j A^{-\top}(Q_j-\Pol_j) ).
	\end{align}
\end{subequations}
	The latter reads as \cite[Eq. 4.32]{monti_feedback_2024}. From \cite[Thm. 4.10]{monti_feedback_2024}, $\buol$ defined in \eqref{eq:def_ol_sequence} is an OL-NE and $\lim_{t\rightarrow \infty} x[t|x_0, \buol(x_0)]=0$ for all $x_0$, thus implying that $\Aol$ is Schur.  \qedd
	\begin{fact}\label{fct:reform_closed_loop_mat} Let $\Pol_i,\Kol_i$ solve \eqref{eq:riccati_open_loop}.  Let $A$ be invertible. Then, \eqref{eq:closed_loop_A} holds true.
	\end{fact}
	\begin{proof} 
		\begin{subequations}  \begin{align}
			\Aol \overset{\eqref{eq:closed_loop_A_explicit}}&{=}A + \tsum_{i\in\mc I}B_i\Kol_i \\
			\overset{\eqref{eq:riccati_open_loop:K}}&{=} A-\tsum_{i\in\mc I} S_i\Pol_i\Aol \\
			&\Leftrightarrow \left(I + \tsum_{i\in\mc I} S_i\Pol_i\right) \Aol = A \label{eq:reform_closed_loop_mat:1} \\
			& \Leftrightarrow \left(I + \tsum_{i\in\mc I} S_i\Pol_i\right) \Aol A^{-1} = I,
		\end{align}
		\end{subequations}
		thus $(I + \tsum_{i\in\mc I} S_i\Pol_i)$ is invertible. Left-multiplying \eqref{eq:reform_closed_loop_mat:1} by $(I + \tsum_{i\in\mc I} S_i\Pol_i)^{-1}$  shows \eqref{eq:closed_loop_A}.
\end{proof}
\subsection{Additional results and proofs of Section \ref{sec:cost_to_go}}\label{app:cost_to_go}
Let us present a preliminary result to the proof of Lemma \ref{le:one_is_lqr}:
\begin{lemma}\label{le:P_is_sum_LQR_tilde}
	Let $(\Pol_i, \Kol_i)_{i\in\mc I}$ solve \eqref{eq:riccati_open_loop} and let Assumption \ref{as:basic_assm:cost}, \ref{as:ol_ne_primitives} hold true. Then, for each $i\in\mc I$, $ \Pol_i = \Plqr_i + \tildeb{P}_i $ where $\tildeb{P}_i$ satisfies
	\begin{equation}\label{eq:p_tilde_sylvester}
		\tildeb{P}_i = (\Alqr_i)^{\top} \tildeb{P}_i \Aol + (\Alqr_i)^{\top} \Plqr_i (\textstyle \sum_{j\neq i} B_j \Kol_j)
	\end{equation}
	and $\Kol_i = \Klqr + \tildeb{K}_i$, where 
	\begin{multline}\label{eq:def_K_tilde}
			\tildeb{K}_i = - (R_i + B_i \Plqr_iB_i)^{-1} \\
			 B_i^{\top}(\tildeb{P}_i\Aol + \Plqr_i(\tsum_{j\neq i} B_j\Kol_j) ).
	\end{multline}
\end{lemma}
\begin{proof}
	{$\Plqr_i$ is symmetric \cite[Thm. 22.2]{hespanha_linear_2018}, thus \eqref{eq:lqr:P} can be written as}
		\begin{align}\label{eq:le:P_is_sum_LQR_tilde:1}
			\Plqr_i = Q_i + (\Alqr_i)^{\top} \Plqr_i A.
		\end{align}
		Now note:
		\begin{align}\label{eq:le:P_is_sum_LQR_tilde:2}
			\begin{split}
				 \Alqr_i \Plqr_i B_i\Kol_i \overset{\eqref{eq:riccati_open_loop:K}}&{=} - \Alqr_i \Plqr_i B_i  R_i^{-1} B_i^{\top}\Pol_i \Aol \\
				\overset{\eqref{eq:lqr:K}}&{=} (B_i\Klqr_i)^{\top} \Pol_i \Aol.
			\end{split}
		\end{align}
		Substituting $ \Pol_i = \Plqr_i + \tildeb{P}_i $ in \eqref{eq:riccati_open_loop:P}: 
		\begin{align*}
			\Plqr_i + \tildeb{P}_i &= Q_i + A^{\top} (\Plqr_i + \tildeb{P}_i)\Aol\\
			 \overset{\eqref{eq:A_cl_lqr}}&{=} Q_i + (\Alqr_i)^{\top} (\Plqr_i + \tildeb{P}_i)\Aol  \\
			& \quad- (B_i\Klqr_i)^{\top} \Pol_i \Aol \\
			\overset{\eqref{eq:le:P_is_sum_LQR_tilde:2}}&{=}  Q_i +  (\Alqr_i)^{\top} (\Plqr_i + \tildeb{P}_i)\Aol \\
			&\quad - \Alqr_i \Plqr_i B_i\Kol_i.
		\end{align*}
		Subtract \eqref{eq:le:P_is_sum_LQR_tilde:1} from the latter to obtain \eqref{eq:p_tilde_sylvester}. Let us rewrite \eqref{eq:riccati_open_loop:K} and \eqref{eq:lqr:K} respectively as:
		\begin{align}
		&\begin{multlined}[t](R_i + B_i^{\top} \Pol_i B_i)\Kol_i =\\ \qquad - B_i^{\top} \Pol_i(A + \tsum_{j\neq i} B_j \Kol_j);\label{eq:K_ol_reform} 
			\end{multlined}\\
		&(R_i + B_i^{\top} \Plqr_i B_i)\Klqr_i = - B_i^{\top} \Plqr_iA. \label{eq:K_lqr_reform}
		\end{align}
		Substitute $ \Pol_i = \Plqr_i + \tildeb{P}_i $ in \eqref{eq:K_ol_reform}:
		\begin{multline*}
			(R_i + B_i^{\top} \Plqr_i B_i)\Kol_i = \\
			 - B_i^{\top} (\Pol_i(A + \tsum_{j\neq i} B_j \Kol_j) + \tildeb{P}_iB_i\Kol_i ).
		\end{multline*}
		which reads as \eqref{eq:def_K_tilde} after subtracting \eqref{eq:K_lqr_reform}.
		\end{proof}

	\paragraph*{Proof of Lemma \ref{le:one_is_lqr}}
		By the Schur property of $\Aol$, it follows that any control action that stabilizes $(A,B_i)$ is stabilizing for $(\widehat{A}_i, \widehat{B}_i)$ for all $i$, thus the system in \eqref{eq:auxiliary_lti_system} is stabilizable. {Let
		 $$\widehat{C}_i := \begin{bmatrix}
			C_i & 0
		\end{bmatrix}.$$ 
		where $C_i$ is as in Assumption \ref{as:ol_ne_primitives}. Clearly, $\widehat{Q}_i = \widehat{C}_i^{\top}\widehat{C}_i$. }
		Let $\widehat{A}_iz = \lambda z$ for some $\lambda\geq1$ and $\widehat{C}_iz=0$, that is, let $z$ an unstable unobservable mode of $\widehat{A}_i$. From the stability of $\Aol$, it has to be $z =[x^{\top}, {0}^{\top}]^{\top}$ for some $x\in\R^{n}$. Then, $\widehat{A}_iz = Ax$ and it has to be $x=0$ by the detectability of $(A, C_i)$. Consequently, $(\widehat{A}_i, \widehat{C}_i)$ is detectable. Following \cite[Cor. 13.8]{zhou_robust_1996}, \eqref{eq:riccati_expanded:P} admits a unique positive semidefinite solution and the corresponding controller is stabilizing. Consider the partition
		\begin{equation*}
			\widehat{P}_i = \begin{bmatrix}
				\widehat{P}_i^{1,1} & \widehat{P}_i^{1,2} \\
				\widehat{P}_i^{2,1} & \widehat{P}_i^{2,2}
			\end{bmatrix}; \quad \widehat{K}_i = [\widehat{K}_i^{1}, \widehat{K}_i^2].
		\end{equation*}
		By expanding \eqref{eq:riccati_expanded} for the blocks $\widehat{P}_i^{1,1}$, $\widehat{P}_i^{2,1}$ and $\widehat{K}_i^1$, one obtains via straightforward calculations:
		\begin{subequations}
			\begin{align}
				\widehat{P}_i^{1,1} &= Q_i + A^{\top}  \widehat{P}_i^{1,1} (A + B_i \widehat{K}_i^1) \label{eq:riccati_blocks_11}\\
				\widehat{K}_i^1 &=  -(R_i+B_i^{\top}\widehat{P}_i^{1,1}B_i)^{-1}B_i^{\top}\widehat{P}_i^{1,1}A \label{eq:riccati_blocks_K_1}\\
				\widehat{P}_i^{2,1} &= (\tsum_{j\neq i} B_j\Kol_j)^{\top} \widehat{P}_i^{1,1}(A + B_i \widehat{K}_i^1) + \nonumber \\
				&(\Aol)^{\top} \widehat{P}_i^{2,1} (A + B_i \widehat{K}_i^1).  \label{eq:riccati_blocks_21}
			\end{align}
		\end{subequations}
		We note that \eqref{eq:riccati_blocks_11} and \eqref{eq:riccati_blocks_K_1} have the same expression as \eqref{eq:lqr:P} and \eqref{eq:lqr:K}, respectively. From $\widehat{P}_i\succcurlyeq 0$, $\widehat{P}_i^{1,1}\succcurlyeq 0$. Thus, $\widehat{P}_i^{1,1}$ is the unique positive semidefinite solution of \eqref{eq:lqr:P} and $\widehat{P}_i^{1,1} = \Plqr_i$, $\widehat{K}_i^1 = \Klqr_i$. Thus, by substituting $\Plqr_i$ and $\Klqr_i$ in \eqref{eq:riccati_blocks_21}, one obtains that $(\widehat{P}_i^{2,1})^{\top}$ must satisfy \eqref{eq:p_tilde_sylvester}. As \eqref{eq:p_tilde_sylvester} is a Stein equation in $\tildeb{P}_i$, its solution is unique following the Schur property of $\Alqr_i$ and $\Aol$ \cite[Lemma 2.1]{jiang_solutions_2003}. Thus, $(\widehat{P}_i^{2,1})^{\top}=\tildeb{P}_i$.\qedd
	\subsection{Proofs of Section \ref{sec:rec_hor_olNE} and \ref{sec:vi_one}}
	\label{app:rec-hor-one}
	\paragraph*{Proof of Theorem \ref{thm:fin_hor_ne_is_inf_hor_ne}}  By the invariance of $\Xol$, $\buex$ is feasible. To prove it is an OL-NE trajectory, we proceed by contradiction. Assume there exists $v_i\in\mc U_{\infty,i}(x_0, \buex_{-i})$ such that
		\begin{equation}\label{eq:le:fin_hor_ne_is_inf_hor_ne:1}
			J_i^{\infty}(v_i|x_0, \buex_{-i})< J_i^{\infty}(\uex_{i}|x_0, \buex_{-i})
		\end{equation} 
		{where $J_i^{\infty}$ is defined in \eqref{eq:inf_hor_objective}.} 
		Let us substitute the definition \eqref{eq:def_u_star_inf_ol} of $\buex$  into \eqref{eq:inf_hor_objective}:
		\begin{align*}
			&J_i^{\infty}(\uex_{i}|x_0, \buex_{-i})  \\
			& = J_i^{\infty}\left(\uol_{i}(\xfh[T]) | \xfh[T], \buol_{-i}(\xfh[T])\right) + \\
			& \quad \tsum_{t=0}^{T-1} \ell_i(\xfh[t], \ufh_{i}[t]) \\
			 \overset{\eqref{eq:V_is_gne_cost_to_go}}&{=}V_i(\xfh[T], \xfh[T]) + \tsum_{t=0}^{T-1} \ell_i(\xfh[t], \ufh_{i}[t]) \\ 
			 \overset{\eqref{eq:cost_rec_hor_ol}}&{=} J_i\left(\ufh_{i}|x_0, \bufh\right).
		\end{align*}
		Denote the state sequence $x_v = \phi(x_0, v_i, \buex_{-i})$: from the definition of \eqref{eq:inf_hor_objective},
		\begin{align*}
			& J_i^{\infty}(v_i|x_0, \buex_{-i}) = J_i^{\infty}(v_i| x_{v}[T], \buol_{-i}(\xfh[T])) +   \\
			& \quad \tsum_{t=0}^{T-1} \ell_i(x_{v}[t], v_{i}[t])  \\
			& \overset{\eqref{eq:V_optimal_xy}}{\geq} V_i(x_{v}[T], \xfh[T]) +  \tsum_{t=0}^{T-1} \ell_i(x_{v}[t], v_{i}[t]) \\
			& \overset{\eqref{eq:cost_rec_hor_ol}}{=} J_i(v_{i}|x_0, \bufh).
		\end{align*}
		Thus, \eqref{eq:le:fin_hor_ne_is_inf_hor_ne:1} contradicts the definition of $\bufh$ in  \eqref{eq:def_u_star}. It follows that $\buex$ is an OL-NE for all $x_0\in\mc X$.\qedd
	\paragraph*{Proof of Lemma \ref{le:shifted_sequence_is_solution}}
		{We first note that, for any $x_0$, by evaluating \eqref{eq:V_value_function} at the optimal input $u_i = (\Klqr_i+\tildeb{K}_i)x_0$ for a generic $i\in\mc I$ and substituting $\Klqr_i+\tildeb{K}_i = \Kol_i$ (Lemma \ref{le:P_is_sum_LQR_tilde}), we obtain}
	\begin{align}\label{eq:evolution_V_nominal}
		\begin{split}
			&	V_i(x_0,x_0) = \ell_i(x_0,\Kol_ix_0)  \\
			&+ V_i\large(Ax_0 +\tsum_{j\in\mc I} B_j \Kol_j x_0,\Aol x_0 \large) \\
			& = \ell_i(x_0,\Kol_ix_0) + V_i(\Aol x_0,\Aol x_0).
		\end{split}
	\end{align}
	We proceed by contradiction and thus assume for some $v_i\in\mc U_{i,T}(\xfh[1], \bush_{-i})$
		\begin{equation}\label{eq:le:switched_is_finite_one:contradiction_assm}
			J_i(v_i|\xfh[1], \bush) < J_i(\ush_{i}|\xfh[1], \bush),
		\end{equation}
	{where $J_i$ is defined in \eqref{eq:cost_rec_hor_ol}.} Denote $x_{v} = \phi(\xfh[1], v_i, \bush_{-i})$.
		Define the auxiliary sequence
		\begin{equation}
			\label{eq:def_v_hat}
			\hat{v}_i[t]:= \begin{cases}
				\ufh_i[t] & \text{if}~ t =0 \\
				{v}_i[t-1] & \text{if} ~ t\in\{1,...,T-1\}.
			\end{cases}
		\end{equation}
		 One can easily verify that 
		\begin{equation}\label{eq:state_evol_v_hat}
			\phi[t+1|x_0, \hat{v}_i, \bufh_{-i}] = x_{v}[t], ~\forall t\in\mc T.
		\end{equation}
		As $v_i\in\mc U_{i,T}(\xfh[1], \bush_{-i})$, clearly $\hat{v}_i$ is also feasible, that is, $\hat{v}_i\in\mc U_{i,T}(x_0, \bufh_{-i})$. By expanding \eqref{eq:cost_rec_hor_ol},
		\begin{align}\label{eq:le:shifted_sequence_is_solution:1}
			\begin{split}
				& J_i(v_i|\xfh[1], \bush)  \\
				&{=V_i(x_{v}[T], \phi[T|\xfh[1],\bush]) + \tsum_{t=0}^{T-1}\ell_i(x_{v}[t], v_i[t]) } \\
				\overset{\eqref{eq:shifted_sol}}&{=}V_i(x_{v}[T], \Aol \xfh[T]) + \tsum_{t=0}^{T-1}\ell_i(x_{v}[t], v_i[t])  \\
				\overset{\eqref{eq:V_value_function}}&{\geq}   V_i(x_{v}[T-1], \xfh[T]) + \tsum_{t=0}^{T-2} \ell_i(x_{v}[t], v_i[t])  \\
				\overset{\eqref{eq:state_evol_v_hat}}&{=}V_i(\phi[T|x_0, \hat{v}_i, \bufh_{-i}], \xfh[T]) \\
				&\quad + \tsum_{t=1}^{T-1} \ell_i(\phi[t|x_0, \hat{v}_i, \bufh_{-i}], \hat{v}_i[t])  \\
				\overset{\eqref{eq:cost_rec_hor_ol}}&{=} J_i(\hat{v}_i|x_0, \bufh) - \ell_i(x_0, \ufh_{i}[0]).
			\end{split}
		\end{align}
		From \eqref{eq:cost_rec_hor_ol} and the definition of $\bush$ in \eqref{eq:shifted_sol},
		\begin{align}\label{eq:le:shifted_sequence_is_solution:2}
			\begin{split}
				&J_i(\ush_{i}|\xfh[1], \bush) = \tsum_{t=1}^{T-1} \ell_i(\xfh[t], \ufh_{i}[t])\\
				&\quad + \ell_i(\xfh[T], \Kol_i \xfh[T])  \\
				&\quad+ V_i( \Aol \xfh[T],  \Aol \xfh[T])  \\
				 \overset{\eqref{eq:evolution_V_nominal}}&{=}V_i(\xfh[T], \xfh[T])+\tsum_{t=1}^{T-1} \ell_i(\xfh[t], \ufh_{i}[t])  \\
				\overset{\eqref{eq:cost_rec_hor_ol}}&{=}J_i(\ufh_i|x_0, \bufh) - \ell_i(x_0, \ufh_i[0]).
			\end{split}
		\end{align}
		By substituting \eqref{eq:le:shifted_sequence_is_solution:1} and \eqref{eq:le:shifted_sequence_is_solution:2} in \eqref{eq:le:switched_is_finite_one:contradiction_assm}, one obtains
		\begin{equation*}
			J_i(\hat{v}_i|x_0, \bufh) < J_i(\ufh_i|x_0, \bufh)
		\end{equation*}
		which contradicts $\bufh$ being a solution of \eqref{eq:def_u_star}.\qedd
		\paragraph*{Proof of Theorem \ref{thm:stability_closed_loop}}
		{Let $\bar{C}:=\col(C_i)_{i\in\mc I}$, where $C_i$ is defined in Assumption \ref{as:ol_ne_primitives}. Clearly, $(A,\bar{C})$ is detectable by the Hautus lemma and Assumption \ref{as:ol_ne_primitives:stab_det}. $(A,\bar{C})$ is then uniformly input/output-to-state-stable \cite[Def. 2.22]{rawlings_model_2017}, \cite[Prop. 3.3]{cai_inputoutput--state_2008}. Furthermore, there exists $L$ such that $A+L\bar{C}$ is Schur. Then, there exists $P_{\mathrm{L}}$ that solves the Lyapunov equation 
		$$ P_{\mathrm{L}} - (A+ L\bar{C})^{\top} P_{\mathrm{L}}(A+LC) = I.$$
		For some $\gamma_{\text{x}}, \gamma_{\text{y}}, \gamma_{\text{u}} >0$ and for all $x$, it holds that }
		\begin{align}
			\begin{split}\label{eq:ioss-lyap} 
			&\|Ax + \tsum_{i\in\mc I} B_iu_i\|^2_{P_\text{L}} - \|x\|^2_{P_\text{L}}  \\
			& \leq - \gamma_{\text{x}}\|x\|^2 + \gamma_{\text{y}} \|\bar{C}x\|^2 + \gamma_{\text{u}}\|\bu\|^2.
			\end{split}
		\end{align}
	{The proof of the latter inequality can be found in \cite[Sec. 3.2]{cai_inputoutput--state_2008} and it is thus omitted. From \eqref{eq:ioss-lyap},
	\begin{subequations}
		\begin{align}
			&\|Ax + \tsum_{i\in\mc I} B_iu_i\|^2_{P_\text{L}} - \|x\|^2_{P_\text{L}}   \nonumber \\
			& \leq  - \gamma_{\text{x}}\|x\|^2 + \tsum_{i\in\mc I} \gamma_{\text{y}} \large(\|x\|_{Q_i}^2 + \gamma_{\text{u}}\|u_i\|^2) \label{eq:reworking_output_mat}\\
			&\leq - \gamma_{\text{x}}\|x\|^2 + \bar\gamma \tsum_{i\in\mc I} \large(\|x\|_{Q_i}^2 + \|u_i\|_{R_i}^2\large),  \label{eq:equivalence_norms}
		\end{align}
	\end{subequations}}
	{where \eqref{eq:reworking_output_mat} follows from $\|\bar{C}x\|^2= \sum_{i}\|x\|_{Q_i}^2$ and \eqref{eq:equivalence_norms} follows from the equivalence of norms and by taking $\bar{\gamma}$ as the maximum multiplicative constant.} {From the direct application of \cite[Theorem B.53]{rawlings_model_2017}, there exists a continuous $\Lambda$, continuous, increasing, unbounded $\alpha_1, \alpha_2$, and a positive definite $\rho$} such that
	\begin{subequations}
		\begin{align}
			&\alpha_1(\|x\|) \leq \Lambda(x) \leq \alpha_2(\|x\|) \label{eq:lyap_rad_unbounded} \\
			&\Lambda(Ax + \tsum_{i\in\mc I} B_iu_i) - \Lambda(x)\leq  \label{eq:ioss_lyap}  \\
			&\qquad-\rho(\|x\|) + \tsum_{i\in\mc I} \large(\|x\|_{Q_i}^2 + \|u_i\|_{R_i}^2\large). \nonumber
		\end{align}
	\end{subequations}
	Consider the candidate Lyapunov function 
	\begin{equation}\label{eq:lyapunov_function}
		V(x)=\Lambda(x) + \tsum_{i\in\mc I} J_i(\ufh_i(x)|x, \bufh(x)),
	\end{equation}
	where $\bufh(x)$ solves \eqref{eq:def_u_star} for the state $x$. Let {$\xfh = \phi(x, \bufh)$.} {Denote the \emph{shifted sequence} $\bush$} as in \eqref{eq:shifted_sol}, and recall that $\bush$ solves \eqref{eq:def_u_star} for the state $\xfh[1]$ following Lemma \ref{le:shifted_sequence_is_solution} and Assm. \ref{as:select_shifted_seq}. {Furthermore, note that 
	\begin{equation} \label{eq:shifted_seq_state_ev}
		\phi[t|\xfh[1], \bush] = \xfh[t+1]\quad \forall t\in\mc T.
	\end{equation}	
	} Then,
	\begin{align}\label{eq:thm:stability_closed_loop:1}
		\begin{split}
		&V(\xfh[1]) - \Lambda(\xfh[1]) \overset{{\eqref{eq:lyapunov_function}}}{=} \smallsum_{i\in\mc I} J_i(\ush_i|\xfh[1], \bush ) \\
		\overset{\eqref{eq:cost_rec_hor_ol}, \eqref{eq:shifted_sol}}&{=}\smallsum_{i\in\mc I}\Big( V_i(\Aol \xfh[T], \Aol \xfh[T]) \\
		& \quad + \ell_i(\xfh[T], \Kol_i \xfh[T]) \\
		& \quad + \tsum_{t=0}^{T-2} \ell_i(\phi[t|\xfh[1], \bush], \ufh_i[t+1])\Big)\\
	\overset{\eqref{eq:evolution_V_nominal}, \eqref{eq:shifted_seq_state_ev}}&{=} \smallsum_{i\in\mc I} \Big(V_i(\xfh[T], \xfh[T])  + \smallsum_{t=1}^{T-1} \ell_i(\xfh[t], \ufh_i[t]) \Big)\\
		\overset{\eqref{eq:cost_rec_hor_ol}}&{=}\tsum_{i\in\mc I} \Big(J_i(\ufh_i|x, \bufh ) - \ell_i(x, \ufh_i[0])\Big) \\
		\overset{\eqref{eq:lyapunov_function}}&{=} V(x) - \Lambda(x) - \tsum_{i\in\mc I}\Big( \ell_i(x, \ufh_i[0]) \Big).
	\end{split}
	\end{align}
We rearrange \eqref{eq:thm:stability_closed_loop:1} and substitute \eqref{eq:stage_cost} to obtain
	\begin{align}\begin{split}
					&V(\xfh[1]) - V(x) \\
					&{=} \Lambda(\xfh[1]) - \Lambda(x) - \smallsum_{i\in\mc I} \|x\|^2_{Q_i} + \|\ufh_i[0] \|_{R_i}^2 \\
			\overset{\eqref{eq:ioss_lyap}}&{\leq} - \rho(\|x\|).
		\end{split}
	\end{align}
	From the invariance of $\Xol$, $\Aol x\in\Xol$. As $\Aol x$ is the terminal state for $\bush$ with initial state $\xfh[1]$, we obtain from the definition of $\mc X$ that $\mc X$ is an invariant set for the closed-loop system. {We conclude that the closed-loop system is asymptotically stable with region of attraction $\mc X$ \cite[Thm. B.13]{rawlings_model_2017}}\qed
	\paragraph*{Proof of Proposition \ref{prop:one_as_vi}}
For this proof we treat finite sequences as column vectors, that is, $u_i=\col(u_i[t])_{{t\in\mc T}}$.
	Let $\bu^*$ solve $\VI(F(\cdot|x_0), \bs{\mc{U}}_T(x_0) )$. {Consider the matrix $\Gamma_i$, defined in \eqref{eq:matrices_VI}:
	We can rewrite the state evolution as
	\begin{align}\label{eq:prop:one_as_vi:1}
		\begin{split} 
			{\begin{bmatrix} \phi[1|x_0, u_i, \bu^*_{-i}]\\ 
				\vdots \\
				\phi[T|x_0, u_i, \bu^*_{-i}]
			\end{bmatrix}} &= \Theta x_0 + \Gamma_{i} u_i + \smallsum_{j\neq i} \Gamma_{j} u^*_j.
		\end{split}
	\end{align}
	Denote the partitions
	$$\Gamma_i = \begin{bmatrix}\overline{\Gamma}_i \\  \underline{\Gamma}_i \end{bmatrix}, \qquad \Theta = \begin{bmatrix} \overline{\Theta} \\ A^T \end{bmatrix},$$ where $\underline{\Gamma}_i$ is the last block row, that is:}
	$$\underline{\Gamma}_i := \begin{bmatrix}
		A^{T-1} B_i & A^{T-2}B_i& \dots & & B_i
	\end{bmatrix}.$$
	Using \eqref{eq:prop:one_as_vi:1} we find the following expression for the terminal states appearing in \eqref{eq:cost_rec_hor_ol}:
	\begin{align} \label{eq:prop:one_as_vi:2}
		\begin{split}
			\phi[T|x_0, \bu^*] &= A^Tx_0 + \underline{\Gamma}_i u^*_i + \tsum_{j\neq i} \underline{\Gamma}_j u^*_j\\
			\phi[T|x_0, u_i, \bu^*_{-i}] &= A^T x_0+ \underline{\Gamma}_i u_i + \tsum_{j\neq i} \underline{\Gamma}_j u^*_j.
		\end{split}
	\end{align}
	Substituting \eqref{eq:prop:one_as_vi:1} and \eqref{eq:prop:one_as_vi:2} in \eqref{eq:cost_rec_hor_ol}, one obtains with straightforward calculations:
	\begin{align}
		\begin{split} 
			& J_i(u_i| x_0, \bu^*) = u_i^{\top}\underline{\Gamma}_i^{\top}(\tfrac{1}{2}\Plqr_i \underline{\Gamma}_i u_i + \tildeb{P}_i  \underline{\Gamma}_i u^*_i  )  \\
			& ~~+ {u}_i^{\top}\underline{\Gamma}_i^{\top} \Pol_i (A^Tx_0 + \tsum_{j\neq i} \underline{\Gamma}_j u^*_j )  \\
			&~~ + \tfrac{1}{2} {u}_i^{\top}(\bar{R}_{i} + \overline{\Gamma}_i^{\top}(I_{T-1}\otimes{Q}_{i})\overline{\Gamma}_i){u}_i \\
			& ~~+  {u}_i^{\top}\overline{\Gamma}_i^{\top}(I_{T-1}\otimes{Q}_{i})(\overline{\Theta} x_0 + \tsum_{j\neq i}\overline{\Gamma}_j u^*_j) \\
			& ~~ + f(x_0, \bu^*),
		\end{split}
	\end{align}
	where $f$ contains the terms that do not depend on $u_i$. {By deriving the latter with respect to $u_i$,
	\begin{align}\begin{split}
		& \nabla J_i(u_i| x_0, \bu^*) = \underline{\Gamma}_i^{\top} (\Plqr_i \underline{\Gamma}_i u_i + \tildeb{P}_i  \underline{\Gamma}_i u^*_i)  \\
		&~~+ \underline{\Gamma}_i^{\top} \Pol_i (A^Tx_0 + \tsum_{j\neq i} \underline{\Gamma}_j u^*_j )  \\
		&~~+ \big(\bar{R}_i +  \overline{\Gamma}_i^{\top}(I_{T-1}\otimes{Q}_{i})\overline{\Gamma}_i \big) u_i  \\
		&~~+ \overline{\Gamma}_i^{\top}(I_{T-1}\otimes{Q}_{i})(\overline{\Theta} x_0 + \tsum_{j\neq i}\overline{\Gamma}_j u^*_j).
	\end{split}\end{align}
	By evaluating the latter at $u_i^*$ and by substituting $\Plqr_i + \tildeb{P}_i = \Pol_i$ (Lemma \ref{le:one_is_lqr}), it can be verified that
	\begin{equation*}
		F(\bu^*|x_0) = \col\big(\nabla J_i(u^*_i|x_0, \bu^*)\big)_{i\in\mc I}.
	\end{equation*}}
	Furthermore, $J_i(\cdot| x_0, \bu^*)$ is convex for every $x_0$, because $\Plqr_i\succcurlyeq 0$, $Q_i\succcurlyeq0$, $\bar{R}_i\succ 0$. Thus, by the definition of VI solution, for any $u_i\in\mc{U}_{i,T}(x_0, \bu^*_{-i} )$:
	\begin{align*}
		0&\leq F(\bu^*|x_0)^{\top} \left( \col(u_i, \bu^*_{-i}) - \bu^*\right)\\
		& = \nabla J_i(u^*_i|x_0, \bu^*)^{\top} (u_i - u_i^*) \\
		&\leq  J_i(u_i|x_0, \bu^*) - J_i(u^*_i|x_0, \bu^*),
	\end{align*}
	that is, $\bu^*$ solves the problem in \eqref{eq:def_u_star}.
	\qedd
	\subsection{Additional results and proofs of Section \ref{sec:rec_hor_clNE}}
	\label{app:rec_hor_clNE}
	Before proving Theorem \ref{thm:convergence_cl_NE}, let us show that the CL-NE satisfies an optimality principle:
	\begin{lemma}\label{le:bellman_closed_loop}
		Let $(\Pcl_i, \Kcl_i)_{i\in\mc I}$ satisfy \eqref{eq:riccati_CL}. Then, for all $i\in\mc I$,
		\begin{multline}\label{eq:bellman_closed_loop}
				\tfrac{1}{2}\|x\|_{\Pcl_i}^2 = \min_{u_i\in \R^{m}}~ \ell_i(x, u_i) + \tfrac{1}{2}  \| \Acl_{-i}x + B_i u_i\|^2_{\Pcl_i}
		\end{multline}
		and the minimum is achieved by $\Kcl_ix$. 
	\end{lemma}
	\begin{proof}
		Let us rewrite \eqref{eq:riccati_CL:K} as
		\begin{equation*}
			\Kcl_i = - (R_i + B_i^{\top}\Pcl_i B_i)^{-1}B_i^{\top}\Pcl_i\Acl_{-i}.
		\end{equation*}
		By setting the gradient of \eqref{eq:bellman_closed_loop} to $0$ and comparing the resulting equation with the latter, one can see that the minimum is achieved by $\Kcl_i x$. By substituting $u_i=\Kcl_i x$ in \eqref{eq:bellman_closed_loop}, the minimum of \eqref{eq:bellman_closed_loop} is
		 \begin{align*}
				& \|x\|^2_{Q_i} + (\Kcl_ix)^{\top} R_i \Kcl_i x + (\Acl x)^{\top} \Pcl_i \Acl x \overset{\eqref{eq:riccati_CL:K}}{=} \\
				& \|x\|^2_{Q_i} - (\Kcl_ix)^{\top} B_i^{\top}\Pcl_i\Acl x +(\Acl x)^{\top} \Pcl_i \Acl x =\\
				& \|x\|^2_{Q_i} + x^{\top} (\Acl - B_i \Kcl_i )^{\top} \Pcl_i \Acl x \overset{\eqref{eq:riccati_CL:P}}{=} \|x\|_{\Pcl_i}^2.
		\end{align*} 
	\end{proof}
\paragraph*{Proof of Theorem \ref{thm:convergence_cl_NE}}
We rewrite the minimization in \eqref{eq:surrogate_gne:cost} as
\begin{multline}\label{eq:thm:convergence_cl_NE:1}
		\min_{u_i\in\mc S_{T-1}^m} \tfrac{1}{2}\big\{\tsum_{\tau=0}^{T-2} \|x[\tau]\|_{Q_i}^2 + \|u_i[\tau]\|_{R_i}^2  \\
		+\min_{u_i[T-1]\in\R^m} \large(\|x[T-1]\|_{Q_i}^2 + \|u_i[T-1]\|_{R_i}^2 \\
		 +\|x[T]\|_{\Pcl_i}^2\large) \big\}.
\end{multline}
By substituting the constraint \eqref{eq:surrogate_gne:dyn} in \eqref{eq:thm:convergence_cl_NE:1} and by applying Lemma \ref{le:bellman_closed_loop}, the inner minimization in \eqref{eq:thm:convergence_cl_NE:1} is solved by $u_i^*[T-1] = \Kcl_i x[T-1]$. Substituting \eqref{eq:bellman_closed_loop}   in the latter, we obtain
\begin{align*}\label{eq:thm:convergence_cl_NE:2}
	\begin{split}
		&\min_{u_i\in\mc S_{T-1}^m} \tfrac{1}{2}\big(\|x[T-1]\|_{\Pcl_i}^2+\smallsum_{\tau=0}^{T-2} \|x[\tau]\|_{Q_i}^2 + \|u_i[\tau]\|_{R_i}^2 \big).
	\end{split}
\end{align*}
By repeating the reasoning backwards in time and substituting the constraint \eqref{eq:surrogate_gne:dyn_0}, \eqref{eq:surrogate_gne:cost} becomes
\begin{align}
	\begin{split}
 \min_{u_i[0]\in\R^m} \tfrac{1}{2}&\big(\|x[0]\|_{Q_i}^2 + \|u_i[0]\|_{R_i}^2   \\
		 &+\big\|Ax[0] + \tsum_{j\in\mc I} B_j u_j[0]\big\|_{\Pcl_i}^2\big).
	\end{split}
\end{align}
	 If $u_j[0]=\Kcl_jx[0]$ for all $j\in\mc I_{-i}$, the minimum is obtained by $u_i^*[0]= \Kcl_ix[0]$ following Lemma \ref{le:bellman_closed_loop}. Thus, $\bu^*$ is a NE.
\qedd

\bibliography{bibl_game_theory}
\bibliographystyle{ieeetr}


\begin{IEEEbiography}[{\includegraphics[width=1in,height
		=1.25in,clip, keepaspectratio]{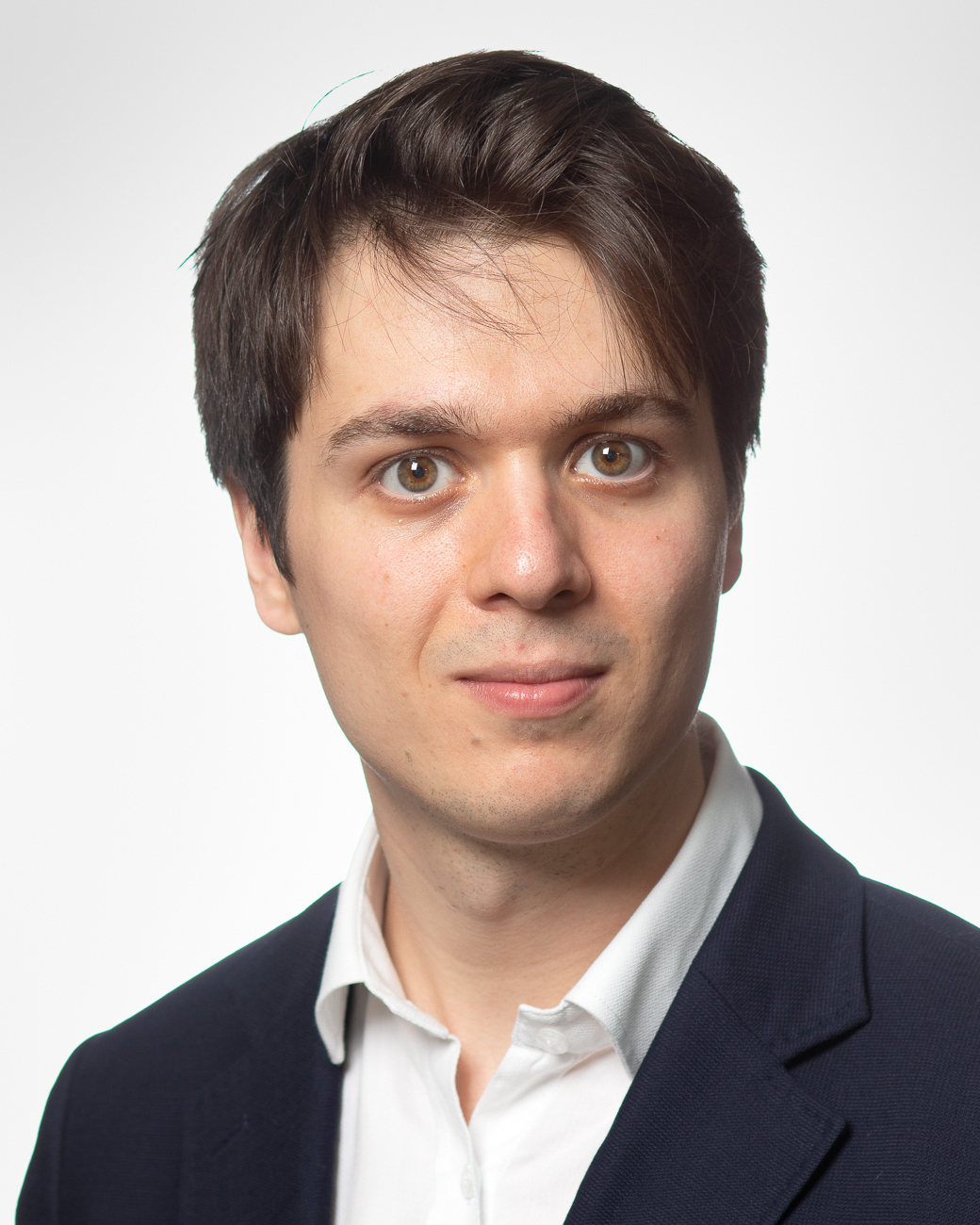}}]{Emilio Benenati} 
	is a post-doctoral researcher at KTH Stockholm, Sweden. He received his Ph.D. degree in 2025 from the Delft Center for Systems and Control, TU Delft, The Netherlands, his Master's degree in 2019 from ETH Z\"urich, Switzerland, and his Bachelor's degree in 2016 from  the University of Catania, Italy. In 2019-2020, he held a research position at the Italian Institute of Technology in Genova, Italy. 
\end{IEEEbiography} 
\vskip -2\baselineskip plus -1fil
\begin{IEEEbiography}[{\includegraphics[width=1in,height
		=1.25in,clip, keepaspectratio]{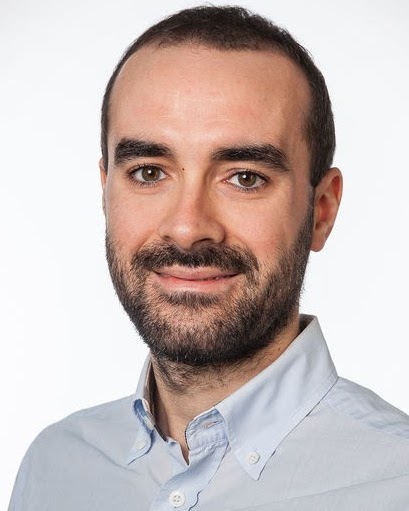}}]{Sergio Grammatico} 
	is an Associate Professor at the Delft Center for Systems and Control, TU Delft, The Netherlands. He received Bachelor, Master and PhD degrees from the University of Pisa, Italy, in 2008, 2009, 2013, respectively. In 2012-2017, he held research positions at UC Santa Barbara, USA, at ETH Zurich, Switzerland, and at TU Eindhoven, The Netherlands. Dr. Grammatico is a recipient of the 2021 Roberto Tempo Best Paper Award. He is an Associate Editor of the IEEE Transactions on Automatic Control and IFAC Automatica.
\end{IEEEbiography}

\end{document}